\documentclass[lettersize,journal]{IEEEtran}

\usepackage{algorithmic}
\usepackage{algorithm}
\usepackage[caption=false,font=normalsize,labelfont=sf,textfont=sf]{subfig}
\usepackage{textcomp}
\usepackage{stfloats}
\usepackage{url}
\usepackage{verbatim}
\usepackage{cite}
\usepackage{amsmath,amssymb,amsfonts}
\usepackage{graphicx}
\usepackage{textcomp}
\usepackage[dvipsnames]{xcolor}
\usepackage{booktabs}
\usepackage{pifont} 
\usepackage{subcaption}
\usepackage[most]{tcolorbox}
\usepackage{tabularx}
\usepackage{array}
\usepackage{threeparttable}
\usepackage{xspace}
\usepackage{multirow}
\usepackage{xcolor}
\usepackage{listings}
\usepackage{orcidlink}
\usepackage{siunitx}
\usepackage{pifont}
\usepackage{makecell}
\tcbuselibrary{breakable}
\usepackage[normalem]{ulem}
\usepackage{siunitx}
\DeclareSIUnit{\pp}{\textup{pp}}

\usepackage{hyperref}
\def\BibTeX{{\rm B\kern-.05em{\sc i\kern-.025em b}\kern-.08em
    T\kern-.1667em\lower.7ex\hbox{E}\kern-.125emX}}

\newcommand{\tracer}{\textsc{Tracer}\xspace}
\newcommand{\corrector}{\textsc{Corrector}\xspace}
\newcommand{\debater}{\textsc{Debater}\xspace}
\newcommand{\gptfive}{\mbox{GPT-5}\xspace}
\newcommand{\gptfour}{\mbox{GPT-4o}\xspace}
\newcommand{\lissafive}{\mbox{LiSSA-5}\xspace}

\definecolor{javablue}{RGB}{0,0,180}
\definecolor{javagreen}{RGB}{0,120,0}
\definecolor{javagray}{RGB}{100,100,100}
\definecolor{javapurple}{RGB}{128,0,128}

\definecolor{lightskyblue}{rgb}{0.53, 0.81, 0.98}
\definecolor{mustard}{rgb}{0.89, 0.72, 0.13}
\definecolor{amber}{rgb}{1.0, 0.49, 0.0}
\definecolor{lavender}{rgb}{0.71, 0.49, 0.86}
\definecolor{carnelian}{rgb}{0.7, 0.11, 0.11}

\newtcolorbox{promptbox}[2]{
  enhanced,
  size=small,
  width=\linewidth,
  fontupper=\footnotesize,
  colback=#2!10,
  colframe=#2!60!white,
  coltitle=black,
  boxrule=0.6pt,
  arc=2pt,
  left=4pt, right=4pt, top=3pt, bottom=3pt,
  title=\textbf{#1}
}

\lstdefinestyle{javaStyle}{
	language=Java,
	basicstyle=\ttfamily\footnotesize,
	keywordstyle=\color{javablue}\bfseries,
	stringstyle=\color{javagreen},
	commentstyle=\color{javagray}\itshape,
	backgroundcolor=\color{gray!5},
	numbers=none,
	frame=none,
	resetmargins=true,
	xleftmargin=0pt,
	showstringspaces=false,
	breaklines=true,
	escapeinside={(*@}{@*)},
	numbers=left, 
	numberstyle=\scriptsize\color{gray}, 
	stepnumber=1,
}

\usepackage{soul} 
\usepackage{xcolor}
\sethlcolor{yellow}

\begin{document}
	
\title{Rethinking Requirement-to-Code Traceability Through 
Single-Agent and Multi-Agent Approaches}

\author{Sarthak 
Banerjee\orcidlink{https://orcid.org/0009-0001-1814-5860},
Sallam 
Abualhaija\orcidlink{https://orcid.org/0000-0001-6095-447X},
~\IEEEmembership{Member,~IEEE,}
Domenico 
Bianculli\orcidlink{https://orcid.org/0000-0002-4854-685X},
~\IEEEmembership{Member,~IEEE,}
\\ and Lionel 
Briand\orcidlink{https://orcid.org/0000-0002-1393-1010},
~\IEEEmembership{Fellow,~IEEE}
}

\markboth{Journal of \LaTeX\ Class Files,~Vol.~14, No.~8, 
August~2021}%
{Shell \MakeLowercase{\textit{et al.}}: A Sample Article Using 
IEEEtran.cls for IEEE Journals}

\maketitle

\begin{abstract}
Requirements-to-code traceability supports essential software engineering activities, including change impact analysis, maintenance, and regulatory compliance. 
Existing approaches typically rely on a retrieval-and-classification pipeline, which suffers from an inherent retrieval bottleneck: relevant source files that are not retrieved cannot be recovered during subsequent analysis. In this paper, we revisit requirements-to-code traceability in light of the substantially larger context windows offered by modern large language models, which make repository-scale analysis feasible without a separate retrieval step. We propose three approaches: a single-agent approach that analyzes the entire code repository in one pass, and two multi-agent approaches based on self-verification and multi-agent debate. Our evaluation shows that the single-agent approach substantially improves recall, achieving 80.9\%, a gain of \SI{38.6}{\pp}  over the state-of-the-art retrieval-based baseline, albeit with a \SI{25.8}{\pp} reduction in precision. 
By leveraging collaborative reasoning, the multi-agent debate-based approach achieves the best overall precision-recall balance, with 24.5\% precision and 71.0\% recall, yielding an F2 score of 49.5\%.  
This represents an improvement of \SI{7.5}{\pp} over the state of the art and \SI{2.7}{\pp} over the single-agent approach. In addition, qualitative analysis provides key insights into highly variable quantitative results: the retrieval approach suits requirements that echo the code’s vocabulary; single-agent, full-context reasoning suits well-documented, consistently named code; and multi-agent debate suits indirect, conflicting, or distributed evidence.
\end{abstract}


\section{Introduction} 
\label{sec:intro}

Establishing trace links between software requirements and source code has long been a persistent challenge in software 
engineering (SE)~\cite{cleland2014}. This challenge becomes even more critical as modern software development increasingly relies on artificial intelligence (AI) coding agents\footnote{The 2026 Stack Overflow survey of 30,000 developers across 169 countries reports that about 66\% of developers use AI coding assistants or agents for software implementation, of whom 73\% indicate daily use--\href{https://stackoverflow.blog/2026/10/06/the-results-of-the-2026-developer-survey-are-here/}{Stack Overflow, \textit{2026 Developer Survey}, October 2026}}.
Today's coding agents can generate entire software projects directly from natural-language (NL) 
requirements~\cite{qian2024}. 
While these technologies offer substantial productivity 
benefits, they also introduce new traceability challenges.

AI-generated code typically lacks explicit trace links that connect the requirements or prompts to their corresponding implementation within the codebase.
The literature has highlighted this gap and emphasized the need for tools to reconstruct these links~\cite {leith2026,velasco2026}.
Furthermore, prior work suggests that AI coding agents cannot yet be relied upon to provide accurate traceability information. For instance, tracing issues to commits through pull requests may be unreliable, because pull request descriptions, often the primary communication channel between AI agents and human developers, do not always accurately reflect the implemented changes~\cite{gong2026}. 
Together, these findings suggest that, although AI-assisted software development can accelerate implementation, it may further weaken the availability and reliability of software traceability information.

Creating and maintaining trace links manually is both time-intensive and error-prone, particularly as systems evolve. 
Whenever code is added, modified, or refactored, the corresponding
trace links must also be updated. In practice, however, these updates are often postponed or omitted altogether. 
As a result, traceability information in industrial projects is frequently missing, incomplete, or outdated~\cite{Veizaga2024}. Recovering these links after development is equally demanding, as it requires examining each requirement against potentially thousands of code artifacts, a task that becomes increasingly costly as repository size grows.

Automated trace link recovery (TLR) is therefore essential, yet remains difficult for several reasons. 
First, a substantial \emph{semantic gap} exists between NL 
requirements and source code, as implementation details are often expressed using terminology that significantly differs from that of the requirement. 
Second, requirements are frequently \emph{scattered} across 
multiple files, meaning that no single file fully captures their 
implementation. Third, the search space is typically 
\emph{highly imbalanced}. In existing requirement-to-code traceability benchmarks~\cite{CoESTDatasets,gao2024}, true trace links often represent only 1\% to 5\% of all possible requirement-code pairs, while the total number of candidate links ranges from roughly \num{6700} to \num{29600} (see Table~\ref{tab:datasets} on page~\pageref{tab:datasets}).
An effective TLR 
approach must therefore identify a small number of correct links among 
thousands of plausible but incorrect candidates.

Recent advances in Large Language Models (LLMs) have led 
to promising progress in automated TLR. Most existing 
approaches follow a retrieval-and-classification pipeline that 
combines Retrieval-Augmented Generation 
(RAG)~\cite{huang2026,ali2024} with LLM-based pairwise 
trace link prediction~\cite{fuchss2025beyond}. Given a 
requirement, these approaches first retrieve the top-$k$ 
semantically similar code files and then ask the LLM to 
determine whether a trace link exists for each requirement-file 
pair.

While this pipeline effectively reduces the search space, it has 
two important limitations. First, it introduces a retrieval bottleneck: any relevant file that is not 
included among the top-$k$ candidates is excluded from 
further consideration. This limitation is particularly problematic 
given the semantic gap between requirements and source 
code, which can prevent relevant files from being ranked 
highly in the first place. Second, evaluating code files independently limits the ability to capture \emph{cross-file traceability}, where a 
requirement is implemented collectively across multiple 
interacting files. Consequently, some trace links may be 
missed even when the relevant files are successfully retrieved.

\begin{figure}
	\centering
	\includegraphics[width=.5\textwidth]{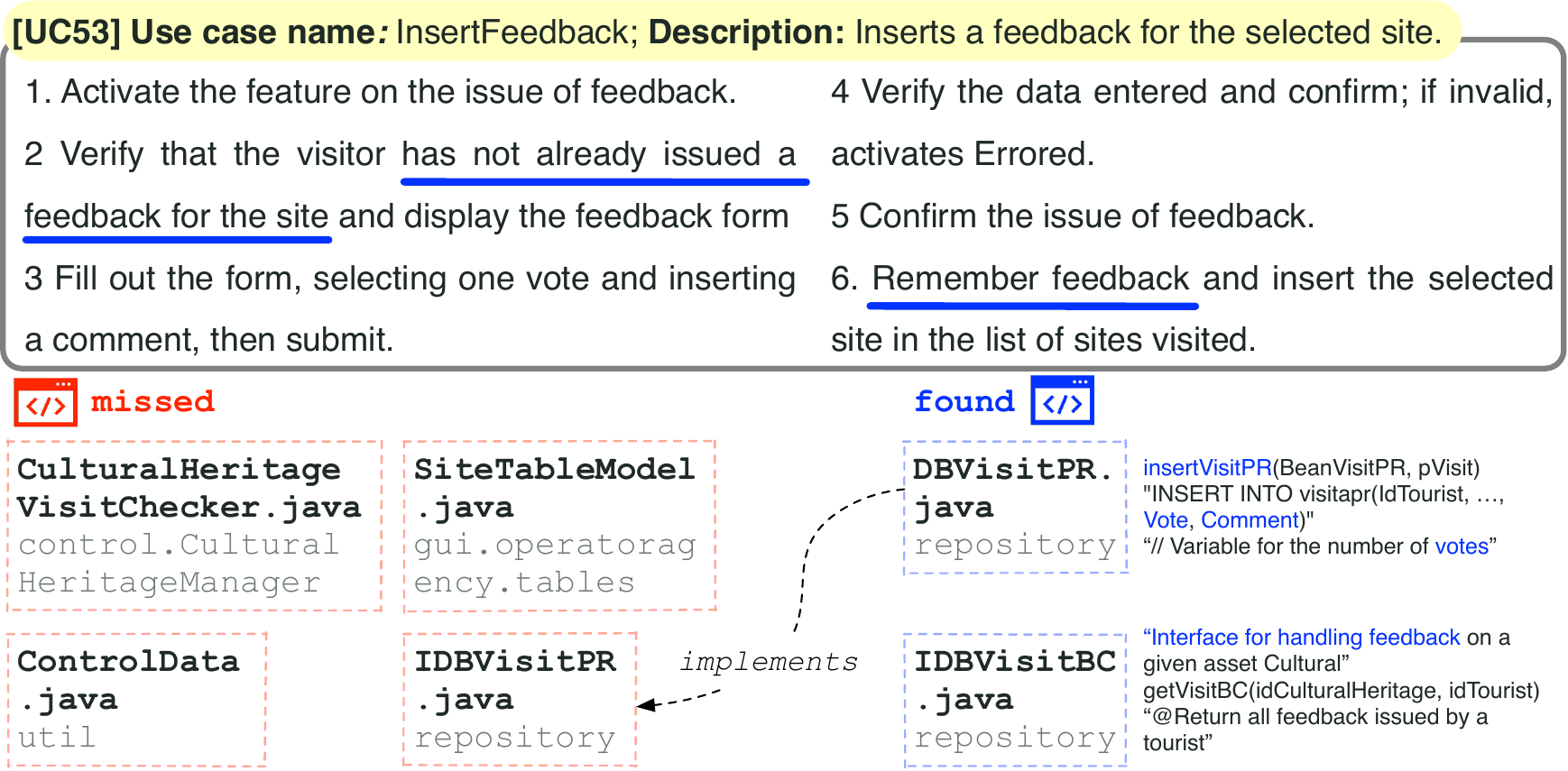}
	\caption{Motivating example: trace links from eTour. }
	\label{fig:example}
	\vspace*{-1em}
\end{figure}

To illustrate these challenges, consider the use case \textbf{UC53} from eTour, a widely used requirements-to-code traceability benchmark~\cite{CoESTDatasets}, illustrated in Fig.~\ref{fig:example}. This use case describes how a tourist submits feedback for a visited site. The system checks that no feedback was issued before, validates the input, stores the feedback, and adds the site to the tourist's list of visited sites. In the benchmark, UC53 is traced to six Java code files: \texttt{IDBVisitBC}, \texttt{DBVisitPR}, \texttt{IDBVisitPR}, \texttt{CulturalHeritageVisitChecker}, \texttt{ControlData}, and \texttt{SiteTableModel}. Existing retrieval-and-classification approaches can identify only two of the six files:  \texttt{IDBVisitBC} and \texttt{DBVisitPR} (marked as \textcolor{blue}{\textbf{found}} in the figure), both of which contain vocabulary close to the use case. The remaining four files (marked as \textcolor{red}{\textbf{missed}} in the figure) are likely missed because the large terminology gap leads to very low textual similarity. Moreover, existing approaches assess each file against the use case in isolation and therefore cannot recover these four files through their code dependencies on the two found ones.

Recent work has attempted to mitigate these issues through 
multi-agent architectures that combine RAG with 
self-verification~\cite{wang2026}. 
However, because self-verification operates on top of the same 
retrieval pipeline, it inherits the retrieval bottleneck: relevant 
files filtered out early cannot be recovered. Moreover, 
self-verification typically reassesses predicted links rather 
than revising them, allowing upstream errors to propagate 
to the final output. Finally, although these approaches employ 
multiple agents, they make limited use of capabilities 
commonly associated with effective multi-agent systems---such 
as collaboration, memory, and coordinated reasoning~\cite{gao2025}.

In this paper, we rethink requirements-to-code traceability by 
asking two questions: (1) Can repository-level reasoning 
recover trace links missed by retrieval-based and pairwise 
approaches? (2) Is a single reasoning pass sufficient? 
To answer the first question, we investigate whether giving an LLM 
access to the entire codebase improves trace recovery, 
particularly for cross-file trace links. To answer the second, we examine the 
role of collaborative reasoning in traceability analysis. Unlike prior self-verification approaches, which primarily reassess predictions, we introduce a multi-agent approach in which agents iteratively critique, refine, and revise candidate trace links.

\noindent\textbf{\textit{Contributions.}} This paper makes the 
following contributions:

\noindent \textbf{(1)} We propose three LLM-based approaches: 
(i) \tracer, a single-agent approach that provides an LLM with 
the entire code repository (rather than a limited set of retrieved 
candidates), enabling repository-level reasoning across 
multiple interacting files to identify cross-file trace links; 
(ii) \corrector and (iii) \debater, two multi-agent approaches that 
build on established self-correction and multi-agent debate 
frameworks~\cite{liang2024,du2024MAD}. We further evaluate 
two prompting strategies: a \textit{conservative} prompt 
($\mathcal{P}_1$), which instructs the model to return only highly confident trace links, and a \textit{comprehensive} prompt 
($\mathcal{P}_2$), which encourages the model to identify all plausible trace links supported by the available evidence, including links that require combining information across multiple artifacts.

\noindent \textbf{(2)} We conduct a comprehensive empirical 
evaluation on three benchmark datasets, namely eTour, iTrust, and Dronology.
Our results show that \tracer with $\mathcal{P}_2$ 
achieves the highest average recall of 80.9\% with an average precision of 17.8\%, and F2 of 46.3\%. 
Compared to the RAG-based approach %
(P = 43.6\%, R = 42.3\%, F2 = 41.5\%), \tracer
improves recall by 38.6 and F2 by 4.8 percentage points (\si{\pp}), at the cost of a \SI{25.8}{\pp} drop in precision. 
Among the multi-agent approaches, \debater with $\mathcal{P}_2$ achieves an average precision of 24.5\%, recall of 71.0\%, and F2 of 49.5\%. Compared to the single agent \tracer, \debater reduces the number of predicted links that an analyst must validate (i.e., confirm or reject) by about 30\% on eTour and iTrust, losing \SI{3.9}{\pp} and \SI{6.3}{\pp} in recall, respectively. On Dronology, it halves the validation effort, but recall drops by \SI{19.3}{\pp}. On all three datasets, \debater also achieves higher precision (24.5\% vs. 17.8\%), i.e., fewer of the links the analyst validates are false positives. 
\debater therefore offers the best trade-off between the human validation effort and the coverage of true links.

\noindent \textbf{(3)} As the quantitative analysis of our experimental results offered limited insights and revealed significant variability across approaches, we complement it with a qualitative analysis of 
all requirements and use cases that have ground-truth links across all datasets. Specifically, 
we manually inspected 57 use cases in eTour, 105 use cases in iTrust, and 57 requirements in Dronology, together with their associated code files. 
Our analysis shows that \tracer excels when trace evidence is 
explicit, whereas \debater is particularly effective at recovering 
trace links supported by indirect, conflicting, or distributed 
implementation evidence. A RAG-based approach only works best when requirements echo the code’s vocabulary. Our findings help practitioners select the most appropriate traceability approach  
based on the characteristics of their artifacts.

\noindent\emph{\textbf{Structure.}} 
Section~\ref{sec:background} presents the necessary 
background. Section~\ref{sec:approach} introduces our 
approaches. Section~\ref{sec:eval} reports on our empirical 
evaluation, and Section~\ref{sec:qualitative} presents the 
qualitative analysis. Section~\ref{sec:related} discusses 
related work, and Section~\ref{sec:conclusion} concludes the 
paper.

\section{Background} 
\label{sec:background}

This section introduces single-agent and multi-agent systems, 
which form the basis of our proposed solutions.

\noindent \emph{\textbf{A single-agent system}} consists of a 
single LLM that is assigned a specific task and is solely 
responsible for generating, refining, and producing the final 
output. Such systems are generally suited to tasks with limited 
complexity and relatively short execution chains, where a single 
model can effectively perform all required reasoning 
steps~\cite{gao2025}. 

LLMs are large-scale transformer-based foundation models 
trained on vast text corpora~\cite{vaswani2017}. Since their 
emergence, these general-purpose models have enabled 
performing a wide variety of tasks through prompting, instead of 
building tools from scratch. For a comprehensive overview of  
prompt engineering, we refer the reader to~\cite{schulhoff2025}. 
A key characteristic of LLMs is their \emph{context window}, 
which defines the maximum amount of information the model can 
process during inference~\cite{Alammar2024}. 
Early generations were constrained by relatively small context 
windows, motivating the development of Retrieval-Augmented 
Generation (RAG), in which relevant information is first retrieved 
and then fed into the model's context~\cite{fan2024}. 
Newer generations (from 2025 onward) have substantially larger 
context windows, enabling more input to be processed 
simultaneously and thereby improving their reasoning 
capabilities~\cite{liu2025}.

\noindent \emph{\textbf{A multi-agent system (MAS)}}  combines two 
or more LLMs that collaboratively perform a single task~\cite{gao2025}. Each LLM is typically assigned a distinct role within the workflow, such as generation, refinement, or validation. This decomposition enables the system to address more complex, 
multi-step tasks and has been shown {to mitigate issues such as \textit{hallucinations} (when LLMs confidently generate false information)} and enhance reasoning capabilities for complex, multi-step tasks~\cite{guo2024}. 

\section{Single-Agent and Multi-Agent Traceability 
Approaches} 
\label{sec:approach}

We propose three LLM-based 
approaches for identifying trace links between requirements and code: a simple single agent approach (\tracer) which benefits from the large context window of modern LLMs, and two more complex multi-agent approaches (\corrector and \debater) implemented based on the self-correction and MAD strategies explained in Section~\ref{sec:background}. 
Addressing the requirements-to-code traceability task requires both identifying relevant candidate code files and validating the resulting trace links. A single-agent approach like \tracer performs only the identification part without careful validation. We therefore propose two multi-agent approaches that explicitly introduce the additional validation step. While \corrector enables correcting the resulting trace links, its reasoning remains limited to the initial trace links. In contrast, \debater enables agents to reason about both selected and rejected trace links through iterative deliberation. Our proposed approaches are motivated by these alternative reasoning paradigms.

\begin{figure*}
	\centering
	\includegraphics[width=\textwidth]{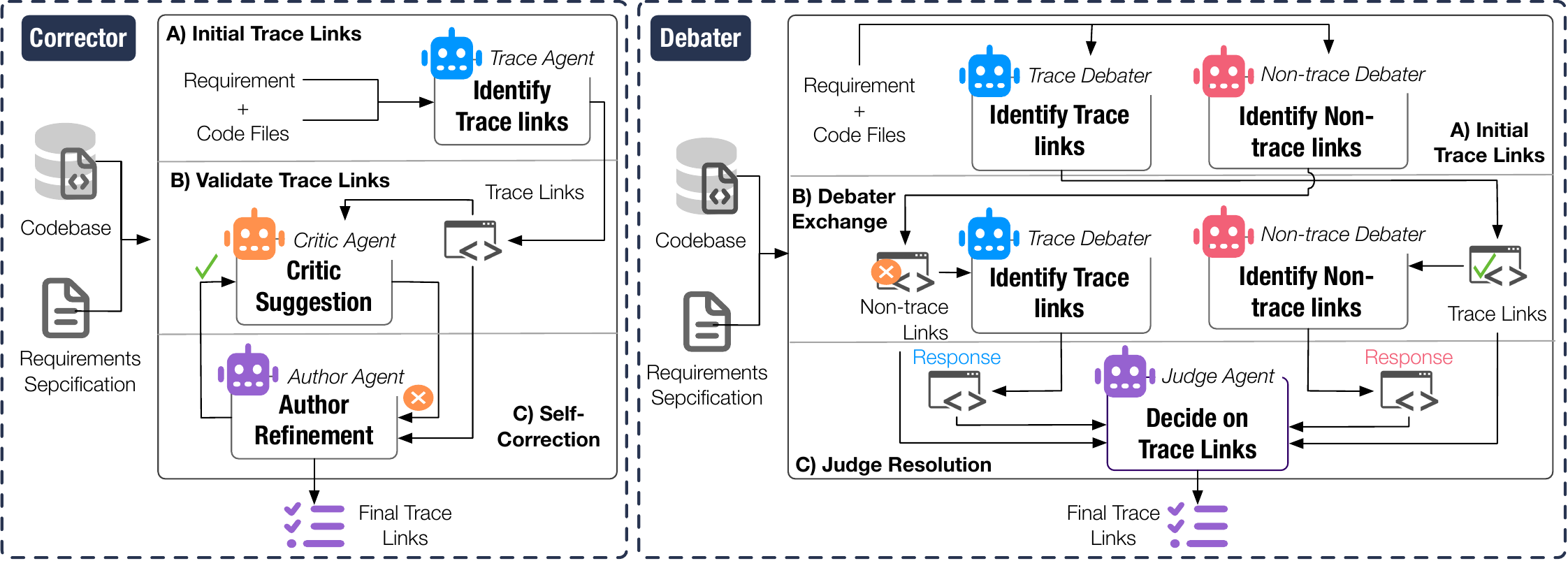}
	\caption{Overview of 
	\corrector (left-side) and	\debater 
	(right-side).}
	\label{fig:solutions}
	\vspace*{-1em}
\end{figure*}

Fig.~\ref{fig:solutions} shows an overview of both \corrector (left-side) and \debater (right-side). We omit \tracer for simplicity.   
Given a requirements specification and a codebase as input, each approach processes a single requirement at a time alongside the complete set of code files. The output for each requirement is then the set of code file names that constitute its trace links.
We describe the approaches next. 

\subsection{\tracer} \label{subsec:tracer} 
In \tracer, a single LLM identifies trace links between a given requirement and all code files in a single pass. Using a customized prompt, the LLM reasons over the requirement and source code artifacts to determine the relevant trace links. 

To keep the focus of this paper on comparing single-agent and multi-agent approaches, we limit the influence of prompt engineering by considering only two prompting strategies. Specifically, we employ two prompt variants designed 
for different traceability objectives. The first, $\mathcal{P}_1$, is a conservative \emph{Chain-of-Thought (CoT)} prompt adapted from existing work~\cite{fuchbeta2025} with adjusted input and output formatting. %
It guides the LLM to derive the trace links by classifying each code file as relevant or irrelevant to the input requirement. 
The second variant, $\mathcal{P}_2$, extends $\mathcal{P}_1$
with explicit instructions that encourage the LLM to generate a broader set of candidate trace links, 
following existing research that favors identifying a comprehensive set of trace links~\cite{etezadi2026}. The prompts $\mathcal{P}_1$ and $\mathcal{P}_2$ are shown in Fig.~\ref{fig:tracer-prompts}.

\renewcommand{\thefootnote}{$\dagger$}

\begin{figure}[t]
  \centering

  \begin{promptbox}{$\mathcal{P}_1$: Conservative Tracing Prompt}{mustard}
    Below are two artifacts from the same software system. Is there a
    traceability link between (1) and (2)?\\
    Give your reasoning, answer with `yes' or `no' and provide a list of
    files.

    \vspace{.3em}
    \textbf{Input}\\
    \texttt{%
      (1) Requirement: \{source\_content\}\\
      (2) Code\_files: \{target\_content\}
    }

    \vspace{.3em}
    \textbf{Output}\\
    \texttt{%
      "trace\_link": "yes or no",\\
      "traced\_files": "if yes list all linked filenames",\\
      "reasoning": "explain the evidence and rationale for these trace links"
    }
  \end{promptbox}

  \begin{promptbox}{$\mathcal{P}_2$: Comprehensive Tracing Prompt}{lightskyblue}
    Below are two artifacts from the same software system. Is there a
    traceability link between (1) and (2)?\\
    \textbf{Maximize recall with traceability links, prefer false positives
    over false negatives.}\\
    Give your reasoning, answer with `yes' or `no' and provide a list of
    files.

    \vspace{.3em}
    \textbf{Input}\\
    same as $\mathcal{P}_1$

    \vspace{.3em}
    \textbf{Output}\\
    same as $\mathcal{P}_1$
  \end{promptbox}
  \vspace{-.5em}
  \caption{Prompts used by \tracer: $\mathcal{P}_1$ (top) and
  $\mathcal{P}_2$ (bottom)  with explicit exploration instructions to identify a broader set of candidate trace links.}
  \label{fig:tracer-prompts}
  \vspace{-1em}
\end{figure}

\subsection{\corrector}
\label{subsec:corrector}
\corrector incorporates self-correction through three distinct agents: 
a \textit{trace agent}, treated as a black-box LLM responsible for initial trace generation; a \textit{critic agent}, which analyzes the quality of the generated 
trace links; and an \textit{author agent}, which refines the trace links based on feedback from the critic. By separating trace generation, critique, and refinement into distinct roles, the approach enables iterative improvement of traceability decisions. 

As shown in Fig.~\ref{fig:solutions}, \corrector consists of 
three steps, split across three stages.  

In stage~A (\textbf{Initial Trace Links}), similar to \tracer, 
a trace agent is instructed to generate an initial set of candidate trace links (step \textbf{identify trace links} in the figure). Compared to \tracer, this step exclusively uses $\mathcal{P}_2$ to ensure the initial set is comprehensive. This choice aligns with the observation that self-correction frameworks are most effective when the initial generation phase prioritizes broad coverage of relevant artifacts~\cite{shinn2023}. By using $\mathcal{P}_2$, the trace agent is encouraged to identify an expansive set of candidate linked artifacts that may be related to the given requirement, even if some of the trace links are weakly supported. 
These weaker trace links are addressed by the other agents through the subsequent self-correction steps (elaborated below). In contrast, using a conservative prompt like $\mathcal{P}_1$ during initial trace generation may prematurely exclude valid trace links that other agents cannot recover. 

Stage~B (\textbf{Validate Trace Links}) consists of a single step, \textbf{Critic Suggestion}, in which the initial set of trace links is passed to the critic agent.  The critic examines the reasoning underlying each link and identifies potential weaknesses, such as weakly supported or questionable traceability relationships. The agent then produces a set of structured suggestions highlighting the identified issues. 

Finally, stage~C (\textbf{Self-correction}) consists of the last step, \textbf{Author Refinement}. Here, the author agent is provided with both the initial trace links and the suggestions generated by the critic agent. Using this information, it revises the trace links to improve the overall quality and consistency of the traceability results. 

The prompts used by the critic and author agents are shown in Fig.~\ref{fig:corrector-prompt}.  
We adopt a simple persona-based prompting strategy for both the critic and author agents. Each agent is assigned a clearly defined role, along with explicit specifications of its expected input and output formats. This design lets each agent focus on its specific responsibilities, reducing the need to optimize prompts to guide its actions.

\begin{figure}[t]
  \centering

  \begin{promptbox}{$\mathcal{P}_3$: Critic Prompt}{amber}
    You are a traceability critic whose job is to identify false positive
    trace links.\\
    A requirement has a trace link only if there is evidence in the provided
    code that it implements or references the requirement.\\
    Do NOT add new files; only evaluate the given code files.

    \vspace{.3em}
    \textbf{Input}\\
    \texttt{%
      (1) Requirement: \{requirement\}\\
      (2) Traced files: \{traced\_files\}\\
      (3) Traced code blocks: \{code\_blocks\}
    }

    \vspace{.3em}
    \textbf{Output}\\
    \texttt{%
      "fp\_files": "List of false positive files",\\
      "reasoning": "Explain why the identified files do not
      implement/reference the requirement"
    }
  \end{promptbox}

  \begin{promptbox}{$\mathcal{P}_4$: Author Prompt}{lavender}
    You are a traceability author whose job is to identify all files that
    implement or reference the requirement.\\
    Add any file where the provided code offers reasonable support.\\
    You are given a set of candidate files and also a list of files that are
    possibly incorrectly traced.

    \vspace{.3em}
    \textbf{Input}\\
    \texttt{%
      (1) Requirement: \{requirement\}\\
      (2) Source code: \{code\_blocks\}\\
      (3) Traced file names: \{traced\_files\}\\
      (4) Possibly false positive traced files: \{fp\_files\}\\
      (5) False positive reasoning: \{fp\_reasoning\}
    }

    \vspace{.3em}
    \textbf{Output}\\
    \texttt{%
      "final\_traced\_files": "List of all files you determine to be valid"
    }
  \end{promptbox}
\vspace{-.5em}
  \caption{Prompts used by \corrector: $\mathcal{P}_3$ (top) flags potentially incorrect trace links and $\mathcal{P}_4$ (bottom) reconsiders the
  initial candidate links given the critic's flags and reasoning.}
  \label{fig:corrector-prompt}
  \vspace{-1em}
\end{figure}

\subsection{\debater} \label{subsec:debater}
\debater consists of three principal agents: a \textit{trace debater}, a \textit{non-trace debater}, and a \textit{judge}. The trace debater is responsible for identifying code files that are likely to be traceable to the given requirement, whereas the non-trace debater focuses on identifying code files that should not be traced to the requirement. Both debaters are provided with the requirement and the 
complete code base in the prompt, similar to the other two approaches. 

As shown in Fig.~\ref{fig:solutions}, \debater operates in three 
stages.
In stage~A, \textbf{Initial Round}, both the trace and non-trace debaters receive an initial prompt that provides a single requirement along with the complete set of code files. The difference is that the trace debater is instructed to identify a set of candidate trace links, while the non-trace debater is instructed to identify the code files that should \textit{not} be traced to the requirement. Both agents are required to provide the reasoning underlying their decisions, either to support the proposed trace links or to justify their exclusion.

In the subsequent stage~B, \textbf{Debater Exchange}, each debater exchanges its outputs from the previous stage. In this cycle, however, the trace debater re-examines the set of non-trace links identified by the non-trace debater and argues for any artifacts that may warrant inclusion as trace links. Conversely, the non-trace debater reviews the trace links proposed by the trace debater and identifies artifacts that should be excluded.

Finally, in stage~C \textbf{Judge Resolution}, the outputs from both the initial round and the subsequent debate exchanges are provided to the judge agent, who then evaluates the competing arguments and resolves any conflicts to produce a final, consolidated set of trace links.

Fig.~\ref{fig:debater-prompts} shows the respective prompts. 
We adopt the prompting strategies and LLM configurations from the requirements engineering literature where MAD was first introduced~\cite{oriol2025}. The trace debater, non-trace debater, and judge agents are all prompted using a simple persona-based approach, in which each agent's role is explicitly defined along with clear 
specifications of the expected input and output formats. This design choice emphasizes each agent's responsibility to perform their assigned tasks without relying on complex prompts.
Furthermore, we restrict the MAD process to a single cycle of debate. This decision allows us to analyze the impact of the MAD framework itself, rather than improvements arising from repeated debate passes.

\renewcommand{\thefootnote}{$\ddagger$}

\begin{figure}[t]
  \centering

  \begin{promptbox}{$\mathcal{P}_5$: Trace \debater Prompt}{lightskyblue}
    You are a debater arguing that the provided files are traceable to the
    requirement.\\
    Return those files that have a trace link.

    \vspace{.3em}
    \textbf{Input}\\
    \texttt{%
      (1) Requirement: \{requirement\}\\
      (2) Files to evaluate: \{non\_trace\_file\_blocks\}
    }

    \vspace{.3em}
    \textbf{Output}\\
    \texttt{%
      "trace\_files": "The files that should be traced to the requirement",\\
      "reasoning": "The reason these files should be traced to the
      requirement"
    }
  \end{promptbox}

  \begin{promptbox}{$\mathcal{P}_6$: Non-Trace \debater Prompt}{carnelian}
    You are a debater arguing that the provided files DO NOT have a trace
    link to the requirement.\\
    Return those files that do not have a trace link.

    \vspace{.3em}
    \textbf{Input}\\
    \texttt{%
      (1) Requirement: \{requirement\}\\
      (2) Files to evaluate: \{trace\_file\_blocks\}
    }

    \vspace{.3em}
    \textbf{Output}\\
    \texttt{%
      "non\_trace\_files": "The files that should NOT be traced to the
      requirement",\\
      "reasoning": "The reason these files should NOT be traced to the
      requirement, give justification based on code"
    }
  \end{promptbox}

  \begin{promptbox}{$\mathcal{P}_7$: Judge Prompt}{lavender}
    You are a moderator. There will be two debaters involved in a debate
    competition. They will present their answers and discuss their
    perspectives (reasonings) on whether a given requirement can be traced
    to some code files.\\
    At the end of each round, you will evaluate their answers and decide
    which files should be traced to the requirement based on the reasoning
    provided.\\
    Take into account all perspectives and choose the right answer.

    \vspace{.3em}
    \textbf{Input}\\
    \texttt{%
      (1) Requirement: \{requirement\}\\
      (2) Trace-files: \{trace\_files\}\\
      (3) Trace-file-reasoning: \{trace\_reason\}\\
      (4) Non-trace-files: \{non\_trace\_files\}\\
      (5) Non-trace-reasoning: \{non\_trace\_reasoning\}\\
      (6) Non-trace-debaters-perspective: \{non\_trace\_debate\}\\
      (7) Trace-debaters-perspective: \{trace\_debate\}
    }

    \vspace{.3em}
    \textbf{Output}\\
    \texttt{%
      "final\_trace\_files": "The files that should be traced to the
      requirement"
    }
  \end{promptbox}
\vspace{-.5em}
  \caption{Prompts used by \debater: $\mathcal{P}_5$ (top) and $\mathcal{P}_6$ (middle) argue for and against candidate trace links and $\mathcal{P}_7$ (bottom) resolves the debate into a final set of trace links.}
  \label{fig:debater-prompts}
  \vspace{-1em}
\end{figure}

\section{Empirical Evaluation} 
\label{sec:eval}

\subsection{Research Questions (RQs)}
\noindent Our evaluation addresses the following RQs:  

\noindent\textit{\textbf{RQ1. How does an LLM-based 
single-agent approach compare to retrieval-based 
approaches for requirement-to-code traceability?}} 
Traditional state-of-the-art methods for TLR 
have utilized RAG specifically to address the context window size limitation 
of available LLMs at the time. This, however, 
restricts the identification of possible genuine trace links to 
what the RAG system retrieves or misses. 
Starting from \gptfive in 2025 onward, LLMs typically support 
large context windows. 
Building on this, RQ1 
compares LiSSA, a state-of-the-art RAG-based approach (introduced in Section~\ref{subsec:baseline}) against \tracer, 
described in Section~\ref{subsec:tracer}.

\noindent\textit{\textbf{RQ2. To what extent does a 
multi-agent LLM-based approach improve 
requirement-to-code traceability compared to a single-agent 
approach?}}
While a single-agent approach benefits from access to a large context window, enabling it to analyze the entire codebase and 
identify candidate trace links in a single generation pass, it has no means of validating or revising the identified links. 
For instance, a single agent that is explicitly instructed to maximize trace link coverage is likely to emit a considerable number of incorrect links. 
By assigning complementary roles to different LLM agents, multi-agent approaches separate the generation of candidate links from their validation. 
As a result, links proposed by one agent can be challenged, revised, or rejected by another, iteratively refining the final set of trace links before it is returned.
In RQ2, we  evaluate two multi-agent approaches: \corrector 
(Section~\ref{subsec:corrector}) 
and \debater (Section~\ref{subsec:debater}). 

\subsection{Baseline}
\label{subsec:baseline}

As a baseline for our study, we consider LiSSA, a 
state-of-the-art RAG-based approach proposed by Fuchs et al.~\cite{fuchbeta2025}. 

LiSSA adopts a two-stage retrieval-and-classification 
architecture. In the first stage, requirements and source code 
artifacts are encoded into a shared embedding space using an  embedding model. Cosine similarity is then used to 
retrieve a set of candidate requirement--code pairs, 
substantially reducing the search space through semantic 
retrieval. This retrieval step serves as the RAG component, 
providing the LLM with a manageable set of potentially 
relevant candidates rather than the complete Cartesian 
product of artifacts. 
In the second stage, the LLM analyzes each retrieved 
requirement--code pair and classifies it as either a valid trace 
link or a non-trace link, guided by a 
conservative prompt 
($\mathcal{P}_1$ in Fig.~\ref{fig:tracer-prompts}). 

Besides evaluating the original LiSSA, we 
developed an independent Python re-implementation 
for two reasons.

First, the replication package released with the 
original LiSSA approach relies on \gptfour as its 
classification model, which supports only a limited 
input context. As discussed earlier, our proposed 
approaches benefit from more recent models with 
larger context windows. Since the LiSSA replication 
package was designed to reproduce the original 
results rather than support customization, replacing 
the underlying LLM would require substantial 
re-engineering. Therefore, re-implementing the 
LiSSA proved to be a more practical option.

Second, the Dronology dataset used in the original 
LiSSA evaluation was not released with the 
replication package. The publicly available version 
used in our study (Section~\ref{subsec:datasets}) 
differs from the one described in the original work in 
the number of requirements. Consequently, 
reproducing the original results cannot be fairly 
compared with those obtained in our study. To keep 
our experimental setting consistent, we 
re-evaluated our re-implementation of LiSSA on the 
same dataset used in our experiments.

Using this re-implementation, we evaluate two LiSSA-based 
configurations: (1) \textit{LiSSA (IR-only)}, executes only the 
retrieval stage of the LiSSA pipeline and returns the Top-$K$ 
most similar source code files for each requirement, thus 
measuring the effectiveness of the embedding-based retrieval 
component in isolation. (2) \textit{LiSSA (\gptfive)}, replaces 
the 
original \gptfour classifier with \gptfive while using the same 
LiSSA pipeline. This variant enables us to assess if advances in 
LLM capabilities affect traceability recovery performance 
when the retrieval stage remains fixed. For both variants, we 
retain the embedding model, cosine-similarity retrieval 
strategy, and retrieval parameter ($K=20$) reported by Fuchs 
et al.~\cite{fuchbeta2025}.

\subsection{Datasets}
\label{subsec:datasets}

To facilitate fair comparison with LiSSA~\cite{fuchbeta2025}, 
we adopt the same benchmark datasets:
eTour, iTrust, and Dronology. These 
datasets are widely used in the traceability literature, making 
them a suitable basis for our empirical investigation.

The eTour and iTrust datasets cover the tourism and 
healthcare domains, respectively, and were originally curated 
by the \emph{Center of Excellence for Software and System Traceability} (CoEST)~\cite{CoESTDatasets}. 
We obtained both datasets directly from the LiSSA replication package to ensure consistency with the original experimental setup, including the number of trace links used for evaluation. 
The Dronology dataset, originally introduced by 
Cleland-Huang et al.~\cite{cleland2018}, covers the aerospace 
domain. Since Dronology is not included in the LiSSA replication package, we use the publicly available version released by Gao et al.~\cite{gao2024}\footnote{\url{https://github.com/huiAlex/TRIAD/tree/main/dataset/dronology}}. 

Table~\ref{tab:datasets} summarizes the characteristics of 
the selected datasets. 
The requirements in eTour and iTrust are represented as use 
cases (UCs), whereas those in Dronology are 
specified as shall-style requirements (RE). All 
datasets contain Java source code files. 
The \emph{CAND} column reports the size of the 
traceability search space, computed as the Cartesian product 
of requirements and code files. Ground-truth trace 
links connect a requirement to one or more code 
files. 

\begin{table}[t] 
	\centering 
	\begin{threeparttable} 
		\caption{Overview of the requirements-to-code datasets 
		used in our study. } 
		\label{tab:datasets} 
		\begin{tabularx}{\columnwidth}{l l c r r r r} 
			\toprule 
			Dataset & Domain & Type & \# Req. & \# Code & 
			CAND\tnote{a} & 	
			TL\tnote{b} \\ 
			\midrule 
			eTour & Tourism & UC & 58 & 116 & 6,728 & 308 \\
			iTrust & Healthcare & UC & 131 & 226 & 29,606 & 286 
			\\ 
			Dronology\tnote{c} & Aerospace & RE & 57 & 184 & 10,488 & 393 \\ 
			\bottomrule 
			\end{tabularx}
			\begin{tablenotes} 
				\footnotesize 				
				\it \item \# Req. = number of requirements; \# Code = 
				number of code files; 
				\item UC, RE: use cases and
				shall-style requirements.
				\item[a] CAND: all possible candidate trace links 
				generated using the Cartesian product of 
				requirements and 
				source code files.
				\item[b] TL: number of ground-truth trace links.
				\item[c] The number of trace links differs from that 
				reported in \cite{fuchbeta2025}. 

			\end{tablenotes} 
		\end{threeparttable} 
		\vspace*{-1em}
\end{table}

\subsection{Evaluation Metrics}
\label{subsec:metrics}

We evaluate all approaches at the trace-link level. 
Given 
the ground-truth trace links provided in each dataset, a true 
positive (TP) denotes a correctly identified trace link. 
A false positive (FP) is a predicted trace link that is not 
present in the ground truth,  while a false negative (FN) 
is a ground-truth trace link that the approach fails to 
recover. 

Following this, we report precision (P),   
recall (R), and F$_\beta$-score, computed as $\mathrm{P}=\mathrm{TP}/(\mathrm{TP}+\mathrm{FP})$, $\mathrm{R}=\mathrm{TP}/(\mathrm{TP}+\mathrm{FN})$, and $\mathrm{F}_\beta=(1+\beta^2)(\mathrm{P}\cdot\mathrm{R})/(\beta^2\cdot \mathrm{P}+\mathrm{R})$ respectively. 

Precision measures the proportion of predicted trace links that are correct, while recall measures the proportion of ground-truth 
trace links successfully recovered by an approach.  
In RE tasks such as traceability, 
recall is generally considered more 
important than precision~\cite{berry2021}. Missing valid trace 
links (i.e., FNs) may reduce the analysts' trust in the automated approach 
and often require a costly manual inspection of the entire 
candidate search space to recover the missed links. We therefore report F2-score (i.e., $\beta=2$), assigning greater weight to recall than precision.

In addition to the above, we report \emph{cost (C)} following Etezadi et 
al.~\cite{etezadi2026b}, computed as C$=\mathrm{TL}/\mathrm{CAND}$, where TL is the number of trace links returned by an  
approach and $\mathrm{CAND}$ is the number of candidate links, i.e., all possible requirements-code pairs (see Table~\ref{tab:datasets}).
Cost estimates the percentage of cost saved by the analyst by reviewing only the trace links produced by the automated approach compared to reviewing all possible candidate links.  
Lower cost values indicate that fewer candidate links must be inspected and, consequently, more manual validation effort is saved.

\subsection{Evaluation Procedure}
\label{subsec:evaluation-procedure}

To answer our RQs, we conduct two experiments (EXPI -- EXPII). 
For each requirement, an approach generates a set of candidate trace links, which are subsequently compared against the ground-truth links. We assess performance using the metrics defined in Section~\ref{subsec:metrics}. We run the experiments on the eTour, iTrust, and Dronology datasets (see Section~\ref{subsec:datasets}). 

\noindent\textbf{\textit{EXPI.}} This experiment addresses 
RQ1. Specifically, we evaluate three variants of the baseline, namely LiSSA (original), LiSSA 
(IR-only), and LiSSA (\gptfive), described in 
Section~\ref{subsec:baseline}. All variants are tested with $\mathcal{P}_1$, the original prompt introduced in the baseline paper~\cite{fuchbeta2025}. 
To assess the effectiveness of the single-agent approach using the large context window, we compare LiSSA with \tracer, both instantiated with \gptfive and tested with $\mathcal{P}_1$. 
To further analyze the impact of a comprehensive prompt, we also evaluate \tracer using $\mathcal{P}_2$. Both prompts are defined in Section~\ref{subsec:tracer}. 

\noindent\textbf{\textit{EXPII.}} This experiment addresses RQ2. 
To assess the impact of incorporating a self-correction mechanism, we compare  \tracer with $\mathcal{P}_2$ against \corrector with $\mathcal{P}_2$. \corrector serves as an upgraded version of \tracer where we introduced another agent dedicated to correcting inherent the errors expected due to maximizing the candidate link coverage in the initial identification phase. 
We then also compare \tracer against a more complex multi-agent approach, \debater, where multiple agents with dedicated roles collaborate to generate the final set of trace links. \tracer and \debater are tested using both prompts.  
We run all approaches using the same underlying LLM, \gptfive and input artifacts. Consequently, any observed differences can be attributed to the reasoning strategy employed by the agents rather than to differences in the candidate search space or model capabilities.

\subsection{Implementation Details} 
\label{subsec:implementation}

All approaches were implemented in Python using 
\emph{LangGraph 1.2.2}~\cite{langgraph2024}, a framework 
for building graph-based LLM applications. LangGraph models 
workflows as directed graphs, enabling a unified 
implementation of both single-agent and multi-agent 
approaches and supporting iterative agent interactions. For all 
experiments, we used \textit{GPT-5-2025-08-07} via the 
OpenAI API~\cite{openaiapi2024}. 
To improve reproducibility, we set the temperature to 0 and left 
all other parameters at their default values. 
For retrieval-based approaches, we applied the same parameters as in the original baseline, namely \textit{OpenAI's text-embedding-3-large} as the embedding model,  $K=20$,  and 
no chunking. 
All experiments were executed on \textit{Ubuntu Linux}.

\subsection{LLM-based Single-agent compared to 
retrieval-based traceability approaches (\textbf{RQ1}) }
\label{subsec:rq1}

\begin{table*}[t]
	\centering
	\begin{threeparttable}
		\caption{Performance of the baseline LiSSA and its two 
		variants (\textbf{RQ1}) }
		\label{tab:lissa_variants}
		\setlength{\tabcolsep}{3pt}
		\begin{tabular}{l ccc ccc ccc ccc}
			\toprule
			& \multicolumn{3}{c}{eTour}
			& \multicolumn{3}{c}{iTrust}
			& \multicolumn{3}{c}{Dronology}
			& \multicolumn{3}{c}{Macro Average}
			\\
			\cmidrule(lr){2-4}
			\cmidrule(lr){5-7}
			\cmidrule(lr){8-10}
			\cmidrule(lr){11-13}
			Variant
			& P ($\Delta$) & R ($\Delta$)   & F2 ($\Delta$) 
			& P ($\Delta$) & R ($\Delta$)   & F2 ($\Delta$)  
			& P ($\Delta$) & R ($\Delta$)   & F2 ($\Delta$)  
			& P & R & F2 
			\\
			\midrule
			
			Original
			& 40.9  & 73.4 & 63.3
			& 19.9 & 45.1 & 36.0
			& 22.6 & 34.4 & 31.1 
			& 27.8 & 51.0 & 43.5 
			\\
			\midrule
			IR-only
			& 21.4 {\scriptsize (-19.5$\downarrow$)} 
			& 80.5 {\scriptsize (+7.1$\uparrow$)}
			& 51.9 {\scriptsize (-11.4$\downarrow$)}
			
			& 5.8 {\scriptsize (-14.1$\downarrow$)}
			& 53.1 {\scriptsize (+8.0$\uparrow$)}
			& 20.2 {\scriptsize (-15.8$\downarrow$)}
			
			& 15.4 {\scriptsize (-7.2$\downarrow$)}
			& 44.8 {\scriptsize (+10.4$\uparrow$)}
			& 32.4 {\scriptsize (-1.3$\uparrow$)}
			
			& 14.2 %
			& 59.5 %
			& 34.8 %
			\\
			
			GPT-5
			& 53.8 {\scriptsize (+12.9$\uparrow$)} 
			& 54.9 {\scriptsize (-18.5$\downarrow$)}
			& 54.7 {\scriptsize (-8.6$\downarrow$)}
			& 31.0 {\scriptsize (+11.1$\uparrow$)}
			& 47.9 {\scriptsize (+2.8$\uparrow$)}
			& 43.2 {\scriptsize (+7.2$\uparrow$)}
			& 46.1 {\scriptsize (+23.5$\uparrow$)}
			& 24.1 {\scriptsize (-10.3$\downarrow$)}
			& 26.7 {\scriptsize (-4.4$\downarrow$)}
			& 43.6 %
			& 42.3 %
			& 41.5 %
			\\
			\bottomrule
		\end{tabular}
		
		\begin{tablenotes}
			\footnotesize
			\item $\Delta$ is the difference in percentage points (pp) 
			with 
			respect to the original LiSSA~\cite{fuchbeta2025}.
		\end{tablenotes}
		
	\end{threeparttable}
\end{table*}

\noindent\textbf{\textit{Retrieval-based Results. }}
Table~\ref{tab:lissa_variants} compares LiSSA and its two variants.
For each variant, the table reports P, R, and 
F2 across the evaluated datasets, together with the relative differences with respect to the original LiSSA approach.

The results show that a retrieval-based approach like LiSSA is 
strongly constrained by its 
retrieval stage, as evidenced by \textit{LiSSA (IR-only)}. On eTour, 
the retriever achieves recall of 80.5\%, recovering 248 trace 
links out of a total of 308 ground-truth trace links (TL in 
Table~\ref{tab:datasets}).  
The 
remaining 
60 links are not included among the Top-$20$ retrieved 
candidates 
and therefore cannot be recovered by the downstream generator. 
This limitation is reflected in the comparably lower recall achieved by both 
LiSSA (original) (73.4\%) and LiSSA (\gptfive) (54.9\%).
The effect is even more noticeable on iTrust and Dronology. 
On iTrust, the retriever recovers only 152 of 286 trace links (R = 53.1\%), missing 134 valid links. On Dronology, it recovers 176 of 
393 trace links (R = 44.8\%), leaving 217 links unrecoverable. These 
results highlight that the retrieval stage establishes a hard upper 
bound on the maximum recall achievable by 
a retrieval-based traceability approach as a whole. 

Replacing \gptfour with \gptfive consistently improves 
precision,  
although its effect on recall varies across datasets. 
On eTour, precision increases from 40.9\% to 53.8\% 
(+\SI{12.9}{\pp}), 
while recall 
decreases from 73.4\% to 54.9\% (-\SI{18.5}{\pp}), resulting in a drop in F2 of \SI{8.6}{\pp}. 
A similar trade-off is observed on Dronology, where precision improves from 22.6\% to 46.1\% (+\SI{23.5}{\pp}) but recall decreases from 34.4\% 
to 24.1\% (-\SI{10.3}{\pp}), reducing F2 by \SI{4.4}{\pp}. In contrast, iTrust benefits from \gptfive in both cases: 
precision increases from 19.9\% to 31.0\% (+\SI{11.1}{\pp}) and recall from 45.1\% to 47.9\% (+\SI{2.8}{\pp}), yielding an F2 improvement of \SI{7.2}{\pp}. 

Since both LiSSA variants operate on the same 
\mbox{Top-$20$} retrieved 
candidates, these differences are attributed to the generator alone. 
Compared with \gptfour, \gptfive appears more conservative, returning fewer trace links and thus achieving substantially higher precision, often at the expense of recall. 

Relative to the original LiSSA configuration, LiSSA with \gptfive achieves an average precision of 43.6\% (+\SI{15.8}{\pp}), recall of 42.3\% (-\SI{8.7}{\pp}), and F2 score of 41.5\% (-\SI{2.0}{\pp}) across all datasets.
Varying the prompt has little impact on overall performance. We tested LiSSA (\gptfive) with the more comprehensive prompt $\mathcal{P}_2$, which achieves an average precision of 30.2\%, recall of 55.1\%, and F2 of 46.6\% across the datasets. While slightly better, retrieval remains the principal bottleneck of the LiSSA pipeline, limiting the recall to a maximum average of 59.5\%.   

As the representative baseline in the remainder of this section, we adopt 
LiSSA (\gptfive) with $\mathcal{P}_1$, referred to as  \textit{\lissafive}, because it combines the original retrieval architecture with the same underlying LLM used to evaluate our approaches.

\noindent\textbf{\textit{Single-agent Results. }}
Table~\ref{tab:rqs} shows the results of \tracer compared to 
\lissafive. 
Using $\mathcal{P}_1$, the same prompt as \lissafive, 
\tracer consistently achieves higher recall across all datasets. 
Improvements in recall range from \SI{1.6}{\pp} on Dronology to \SI{16.4}{\pp} on iTrust. 
Unlike \lissafive, which is restricted to the retrieved 
candidates, \tracer reasons directly over the complete set of source-code files. 
This broader context appears to improve the recovery of valid trace links that would otherwise be discarded during retrieval. 
These gains suggest that eliminating retrieval-based filtering is particularly beneficial when valid trace links are weakly represented by lexical similarity alone, as we further elaborate in the qualitative analysis 
below. 
Precision remains comparable between the two 
approaches, likely because they rely on the same underlying model, i.e., \gptfive.

\begin{table*}[t]
	\centering
	\begin{threeparttable}
		\caption{Performance of retrieval-based and single-agent 
		(\textbf{RQ1}), 
		and multi-agent (\textbf{RQ2})
		approaches. Best results are highlighted in \textbf{bold} for each category.  
		}
		\label{tab:rqs}
		
		\setlength{\tabcolsep}{4pt}
		
		\begin{tabular}{lr cccccc cccccc cccccc ccc}
			\toprule
			
			&& \multicolumn{6}{c}{eTour}
			& \multicolumn{6}{c}{iTrust}
			& \multicolumn{6}{c}{Dronology} 
			& \multicolumn{3}{c}{Macro Average}
			\\
			\cmidrule(lr){3-8}
			\cmidrule(lr){9-14}
			\cmidrule(lr){15-20}
			\cmidrule(lr){21-23}			
			
			Approach & $\mathcal{P}$\tnote{*}
			& TP & FP & FN & P & R & 
			F2
			& TP & FP & FN & P& R & 
			F2
			& TP & FP & FN & P & R & 
			F2
			& P & R & F2 \\
			\midrule
			
			\lissafive &$\mathcal{P}_1$
			& 169 & \textbf{145} & 139 & \textbf{53.8} & 54.9 & 54.7
			& 137 & \textbf{304} & 149 & \textbf{31.0} & 47.9 & 43.2
			& 95 & \textbf{111} & 298 & \textbf{46.1} & 24.1 & 26.7 
			& \textbf{43.6} & 42.3 & 41.5\\

			\midrule
			
			\tracer &$\mathcal{P}_1$
			& 188 & \textbf{148} & 120 & \textbf{56.0} & 61.0 & \textbf{59.9}
			& 184 & \textbf{426} & 102 & \textbf{30.2} & 64.3 & \textbf{52.5}
			& 101 & \textbf{126} & 292 & \textbf{44.5} & 25.7 & 28.1 
			& \textbf{43.5} & 50.4 & \textbf{46.8}  \\
			
			\tracer &$\mathcal{P}_2$
			& \textbf{269} & 872 & \textbf{39} & 23.6 & \textbf{87.3} & 56.7
			& \textbf{245} & 1874 & \textbf{41} & 11.6 & \textbf{85.7} & 37.5
			& \textbf{274} & 1225 & \textbf{119} & 18.3 & \textbf{69.7} & \textbf{44.6} 
            & 17.8 & \textbf{80.9} & 46.3 \\
			\midrule

			\corrector & $\mathcal{P}_2$
			& 239 & 420 & 69 & 36.3 & 77.6 & 63.2
			& 209 & 1117 & 77 & 15.8 & 73.1 & 42.3
			& 192 & 478 & 201 & 28.7 & 48.9 & \textbf{42.8}
			& 26.9 & 66.5 & 49.4 \\
			
			\debater & $\mathcal{P}_1$
			& 217 & \textbf{240} & 91 & \textbf{47.5} & 70.5 & 
			\textbf{64.2}
			& 213 & \textbf{681} & 73 & \textbf{23.8} & 74.4 & 
			\textbf{52.3}
			& 141 & \textbf{249} & 252 & \textbf{36.1} & 35.9 & 
			35.9
			& \textbf{35.8} & 60.2 & \textbf{50.8} \\
			
			\debater & $\mathcal{P}_2$
			& \textbf{257} & 540 & \textbf{51} & 32.2 & 
			\textbf{83.4} & 63.3
			& \textbf{227} & 1286 & \textbf{59} & 15.0 & 
			\textbf{79.4} & 42.7
			& \textbf{198} & 554 & \textbf{195} & 26.3 & 
			\textbf{50.4} & 42.6
			& 24.5 & \textbf{71.0} & 49.5 \\
			\bottomrule
		\end{tabular}
		
		\begin{tablenotes}
			\footnotesize
			\item[*] $\mathcal{P}_1$ and $\mathcal{P}_2$ are the 
			prompts in Fig.~\ref{fig:tracer-prompts}.
		\end{tablenotes}
		\vspace{-1em}
	\end{threeparttable}
\end{table*}

Compared to $\mathcal{P}_1$, the more comprehensive recall-oriented prompt $\mathcal{P}_2$ gives \tracer the highest recall: 87.3\% on eTour, 85.7\% on iTrust, and 69.7\% on Dronology. These results exceed the upper-bound recall obtained by the retriever alone (LiSSA-IR-only) by a large margin, mostly notable on iTrust (+\SI{32.6}{\pp}) and Dronology (+\SI{24.9}{\pp}), 
with a non-negligible improvement on  eTour (+\SI{6.8}{\pp}). 
These improvements indicate that the model is capable of 
identifying substantially more valid trace links when encouraged to prioritize recall. 
The gains, however, come at the expense of precision, as 
the broader exploration of candidate links also increases the number of FPs. 

Table~\ref{tab:cost} shows the increase in cost, i.e., the additional manual human effort required to validate the generated links, in the case of \tracer ($\mathcal{P}_2$), compared to $\mathcal{P}_1$. 
On eTour, cost increases from inspecting 314 trace links 
(corresponding to 4.99\% of the total candidate links 
under~$\mathcal{P}_1$) to 1141 trace links (16.96\% of the 
total 
candidates) in the case of $\mathcal{P}_2$,  
while ensuring that fewer genuine trace links are missed, increasing  
recall from  
61.0\% (missing 120 links) to 87.3\% (missing only 39). 
A similar trend is observed on iTrust, where recall 
improves from 64.3\% to 85.7\% at the expense of a cost increase from 2.06\% to 7.16\%. 
The trade-off is most obvious on Dronology, where recall increases from 25.7\% to 69.7\% but cost 
grows from 2.16\% to 14.29\%. 
Nonetheless, the analyst can have considerably greater confidence in the completeness of the recovered trace links. 

Overall, the additional 
manual 
validation 
effort appears justified by the substantial reduction in missed 
trace links. Specifically, compared to \tracer 
($\mathcal{P}_1$), \tracer ($\mathcal{P}_2$) reduces the 
average proportion of unrecovered trace links from 49.7\% to 
19.1\%, while increasing the average validation cost from 
3.07\% to 12.80\%. Given the importance of recall in 
traceability tasks, this represents a favorable trade-off 
between trace-link completeness and manual inspection effort.

Consequently, \tracer with $\mathcal{P}_2$ is preferred when maximizing trace-link recovery is the 
primary objective, whereas the more conservative approach 
\tracer with $\mathcal{P}_1$
remains a good choice when minimizing 
validation effort is more important.

\begin{table*}[t]
	\centering
	\caption{Trade-off between TLR 
		and manual validation effort for 
		\lissafive and \tracer
		(\textbf{RQ1}), 
		and \debater (\textbf{RQ2})
		 }
	\label{tab:cost}
	
	\begin{tabular}{l cc|cc cc|cc cc|cc}
		\toprule
		
		& \multicolumn{4}{c}{eTour}
		& \multicolumn{4}{c}{iTrust}
		& \multicolumn{4}{c}{Dronology} \\
		
		\cmidrule(lr){2-5}
		\cmidrule(lr){6-9}
		\cmidrule(lr){10-13}
		
		Approach
		& Pred. / CAND & C &  R & Missing 
		& Pred. / CAND & C& R & Missing 
		& Pred. / CAND & C& R & Missing \\
		
		\midrule
		
		\lissafive
		& 314 / 6,728  & 4.67  & 54.9 & 45.1 
		& 441 / 29,606  & 1.49 & 47.9  & 52.1 
		& 206 / 10,488    & 1.96  & 24.1 & 75.9\\

        		\midrule

		\tracer ($\mathcal{P}_1$)
		& 336 / 6,728   & 4.99 & 61.0 & 39.0 
		& 610 / 29,606   & 2.06 & 64.3 & 35.7 
		& 227 / 10,488   & 2.16 & 25.7 & 74.3  \\
		
		\tracer ($\mathcal{P}_2$)
		& 1,141 / 6,728 & 16.96 & 87.3 & 12.7 
		& 2,119 / 29,606 & 7.16& 85.7 & 14.3 
		& 1,499 / 10,488 & 14.29 & 69.7 & 30.3  \\
		
		\midrule
		
		\debater\ ($\mathcal{P}_1$)
		& 457 / 6,728 & 6.79 & 70.5 & 29.5 
		& 894 / 29,606 & 3.02 & 74.4 & 25.6 
		& 390 / 10,488 & 3.72 & 35.9 & 64.1 \\
		
		\debater\ ($\mathcal{P}_2$)
		& 797 / 6,728 & 11.85 & 83.4 & 16.6 
		& 1,513 / 29,606 & 5.11 & 79.0 & 21.0 
		& 752 / 10,488 & 7.17 & 50.4 & 49.6 \\
		\bottomrule
	\end{tabular}
	
	\begin{tablenotes}
		\footnotesize
		\item Pred. = number of trace links predicted by the 
		approach, CAND = all possible candidate trace links, C = 
		validation cost (\% of inspected candidate 
		links), R = 
		recall, and Missing = \% of unrecovered ground-truth 
		trace 
		links.
	\end{tablenotes}
	\vspace{-1em}
\end{table*}

    \begin{tcolorbox}[breakable,enhanced, colback=white, colframe=black, boxrule=0.4pt, arc=0pt, 
    arc=0pt, left=2pt, right=2pt, top=2pt, bottom=2pt, boxsep=2pt, 
    width=0.95\linewidth] 
\textbf{Answer to RQ1.} Reasoning over the entire codebase enables 
\tracer to recover trace links that retrieval-based approaches 
do not consider, improving recall at comparable precision when both use the same prompt.
The comprehensive prompt 
($\mathcal{P}_2$) 
further increases recall beyond the upper bound imposed by retrieval, by allowing the single agent to 
output a broader set of trace links.
These gains, however, come at the cost of substantially more false 
positives and higher manual validation effort. We argue, 
however, that this effort is acceptably small in many contexts.
Since recall is generally prioritized in traceability, and
 links initially missed may remain permanently undiscovered, \tracer ($\mathcal{P}_2$) 
is considered the preferred configuration when completeness is the primary goal, whereas \tracer ($\mathcal{P}_2$) is preferable when validation effort is constrained.

 \end{tcolorbox}

\subsection{Multi-agent compared to single-agent 
traceability approaches (\textbf{RQ2}) }
\label{subsec:rq2}

\noindent\textbf{\textit{Self-correction Results.}} 
Table~\ref{tab:rqs} shows that
\corrector, compared to \tracer, improves the average precision by 
9.1 pp (26.9\% vs. 17.8\%) by substantially reducing FPs, from 
872 to 420 on eTour, 1874 to 1117 on iTrust, and 1225 to 478 on 
Dronology. This reduction lowers manual validation costs.  The results confirm that errors made by \tracer $\mathcal{P}_2$ can indeed be mitigated \emph{after} trace link generation through the self-correction mechanism. 
However, improving precision comes at the expense of recall, which drops from 80.9\% to 66.5\%, indicating that the mechanism can incorrectly filter out some correct trace links. 

While \corrector achieves a slightly higher average F2 score than \tracer (49.4\% vs. 46.3\%), it is still limited to only correcting errors in the generated trace links and not recovering valid links that \tracer ($\mathcal{P}_2$) potentially failed to identify in the first place. 
This limitation makes \corrector less attractive in practice than \tracer because it has a higher computational cost and yields only limited performance gains.

\noindent\textbf{\textit{MAD Results.}}
Table~\ref{tab:rqs} further shows \debater's results. 
Among the evaluated approaches, \debater achieves the 
highest average F2 score overall, regardless of the prompt 
used: 50.8\% with $\mathcal{P}_2$ and 49.5\% with 
$\mathcal{P}_1$. 
Compared to the single-agent 
\tracer with the same prompt, \debater with $\mathcal{P}_1$ loses 
\SI{7.7}{\pp} in precision while improving recall by \SI{9.8}{\pp}. 
In contrast, with 
$\mathcal{P}_2$, it gains \SI{6.7}{\pp} in precision at the cost 
of an 
average loss of  \SI{9.9}{\pp} in recall. 
Compared to \corrector, \debater with $\mathcal{P}_2$ 
achieves, on average, slightly lower precision (24.5\% vs. 
26.9\%), but improves recall by \SI{4.5}{\pp} (71.0\% vs. 
66.5\%). 

From a validation-cost perspective, \debater substantially 
reduces the number of candidate links requiring manual review. 
As shown in Table~\ref{tab:cost}, compared to \tracer 
with 
$\mathcal{P}_2$, \debater ($\mathcal{P}_1$) predicts fewer 
links, but at the cost of missing a larger proportion of ground 
truth links, ranging  from $\approx$26\% on iTrust to 64\% on 
Dronology. 
\debater ($\mathcal{P}_2$) also substantially reduces the 
proportion of candidate links to review, from about 17\% to 12\% 
on eTour, from 7\% to 5\% on iTrust, and from 14\% to 7\% on 
Dronology, relative to \tracer. This reduction likewise leads to 
missing true links, although to a lesser extent than \debater 
($\mathcal{P}_1$), ranging from $\approx$17\% on eTour to 
50\% for Dronology.

Since our evaluation prioritizes recall, we select \debater ($\mathcal{P}_2$) 
over \debater ($\mathcal{P}_1$), despite it achieving slightly lower 
macro-average F2 scores.  
\debater ($\mathcal{P}_2$) improves macro-average recall by \SI{10.8}{pp} 
(71.0\% vs.\ 60.2\%), whereas \debater ($\mathcal{P}_1$) leads in 
macro-average F2 by only \SI{1.3}{pp} (50.8\% vs.\ 49.5\%). In other words, 
\debater ($\mathcal{P}_2$) delivers a substantial gain in recall at a 
marginal cost in F2.

	\begin{tcolorbox}[breakable,enhanced, colback=white, colframe=black, boxrule=0.4pt, arc=0pt, 
    arc=0pt, left=2pt, right=2pt, top=2pt, bottom=2pt, boxsep=2pt, 
    width=0.95\linewidth]

\textbf{Answer to RQ2.} Explicit reasoning about both trace and non-trace links through multi-agent debate proves effective. Compared to \tracer, \debater increases precision and reduces the number of links to validate, at some loss in recall that is most pronounced on Dronology. 
\debater ($\mathcal{P}_2$) yields the best trade-off between link coverage and validation effort, with a macro-average recall of 71\% and validation costs ranging from 5.11\% (iTrust) to 11.85\% (eTour).
\end{tcolorbox}

\subsection{Threats to validity}
\label{sec:threats}

In this section, we discuss the threats to validity of our study,
following the guidelines for empirical studies in software
engineering involving LLMs~\cite{baltes2026}.

\noindent\textbf{\textit{Internal Validity.}}
A potential threat is the sensitivity of the approaches to the prompt formulation. To reduce this effect, we adopted existing prompts and applied them consistently across the different approaches.

\noindent\textbf{\textit{External Validity.}}
The main external threat concerns the generalizability of our 
findings across models and software projects. 
To partially mitigate this threat, we evaluate our approaches on 
different datasets spanning multiple domains and requirement types. 
However, all evaluated systems are implemented in Java. 
Further studies are therefore needed to improve the generalizability of our findings to other 
programming languages and larger software systems.

\noindent\textbf{\textit{Construct Validity.}}
Following prior work~\cite{fuchbeta2025}, we use precision, 
recall, and F2 at the trace-link level. We additionally report 
validation cost, measured as the percentage of candidate links 
requiring manual inspection, although this assumes equal 
inspection effort across all links and should therefore be interpreted as an approximation of reviewer workload.

\noindent\textit{\textbf{Conclusion Validity.}} Our experiments rely primarily on \gptfive. As LLMs are inherently non-deterministic, repeated executions may yield slightly different outputs. To improve reproducibility, we set the temperature to 0, document all prompts and generated responses, and provide the exact model versions used in our work.
Nevertheless, reproducibility remains limited by the availability of information about commercial LLMs, whose proprietary components and evolution are not fully disclosed.

\renewcommand{\thefootnote}{\arabic{footnote}}

\section{Qualitative Analysis}
\label{sec:qualitative}
Complementary to the quantitative results presented above, 
we conduct a 
qualitative analysis based on a manual inspection of 57 use cases in eTour, 105 use cases in iTrust, and 57 requirements in Dronology\footnote{Note that not all requirements reported in Table~\ref{tab:datasets} had ground-truth links.}, together with their corresponding code files connected through true trace links.
We further manually inspect the ground-truth trace links and the trace links produced by the evaluated approaches to better understand their strengths, limitations, and underlying behaviors.

Our analysis is informed by a recent study on the 
relationship between requirements quality and 
TLR~\cite{tobias2026}. 
In brief, several requirement qualities that are often considered 
defects in the RE literature~\cite{frattini2022} can actually 
benefit traceability. 
These include \textit{inconsistent levels of abstraction}, which 
expose implementation-oriented details; \textit{alternatives 
blended 
	into the main flow}; and \textit{tangled requirements}, where 
	a single 
use case is implemented across multiple code files. 
Such characteristics introduce implementation-level 
vocabulary into 
requirements, thereby narrowing the semantic gap between 
requirements and source code and making traceability links 
easier to 
recover. 

The study further reports that certain requirements quality 
issues 
can hinder traceability 
recovery. These include requirements that \textit{start 
	without a clear noun phrase}, contain \textit{anaphoric 
	references}, 
as well as 
\textit{scattered requirements}, where the implementation of a 
single 
functionality is spread 
across multiple use cases. These characteristics 
obscure or fragment information in the use case, increasing 
the 
semantic gap between requirements and source code.

While requirements quality is an important factor in traceability 
recovery, we argue that the characteristics of the codebase 
are 
equally important, as requirements-to-code traceability 
depends on 
both artifact types. %
Our qualitative analysis examines both requirements vis-à-vis their associated code 
files to 
better understand why some trace links are recovered while 
others 
are missed. 

    \subsection{Qualitative Comparison of Retrieval-Based 
and Single-Agent Approaches}

\vspace*{.3em}
\noindent\fbox{\textbf{eTour}}
Consistent with the positive traceability factors discussed above, 
the eTour use cases follow a structured template and frequently 
incorporate implementation-oriented vocabulary, including 
alternative flows. 
These characteristics reduce the semantic gap between 
requirements and source code by increasing lexical and behavioral 
overlap. This is reflected in the performance of LiSSA (IR-only), 
which achieves a recall of 80.5\% (Table~\ref{tab:lissa_variants}). 
We therefore attribute the strong retrieval performance primarily to 
the close alignment between requirement descriptions and 
implementation artifacts.

\begin{tcolorbox}[ 
	breakable, 
	colback=gray!10, 
	colframe=black!30, 
	left=1mm, right=1mm, 
	top=0.5mm, bottom=0.5mm, 
	boxsep=1mm, 
	arc=1mm,
	fonttitle=\bfseries,
	title={eTour UC$53$: ``InsertFeedback''},  
	] 
\small
\label{example:eTour_UC53}
\textbf{Use case name:} InsertFeedback\\
\textbf{Description:} Inserts a feedback for the selected site.\\
\textbf{Participating actor:} initialized by Tourist\\
\textbf{Entry conditions:} The tourist card is in a particular site.\\
\textbf{Flow of events User System:} \\
1. Activate the feature on the issue of feedback.\\
2. Verify that the visitor has not already issued a feedback for the
site and displays a form for entering the feedback for the selected
site. In case the visitor has already issued a feedback for the site
selected, activates the use case FeedbackAlreadyReleased.\\
3. Fill out the form, selecting one vote and inserting a comment, then
submit.\\
4. Verify the data entered and confirm the cheide. Where the data
is invalid or insufficient, the system activates the use case
Errored. \\
5. Confirming the issue of feedback. \\
6. Remember feedback and inserts the selected site in the list of
sites visited. \\
\textbf{Exit conditions:} The system shall notify the successful
combination of feedback to the site. The Tourist cancel the operation.
Interruption of the connection to the server ETOUR.
\end{tcolorbox}

\begin{figure}[t] \centering \footnotesize

	\begin{tabularx}{\linewidth}{
			>{\raggedright\arraybackslash}X c 
			>{\raggedright\arraybackslash}X}
		\toprule
		\textbf{Ground Truth} & & \textbf{Predicted} \\
		\midrule
		
		\texttt{IDBVisitBC.java} & TP&
		\texttt{IDBVisitBC.java} \\
		\midrule
		\texttt{DBVisitPR.java} & TP &
		\texttt{DBVisitPR.java} \\

		\midrule
		\texttt{IDBVisitPR.java} & TP &
		\texttt{IDBVisitPR.java} \\
\midrule
		
		\makecell[l]{\texttt{CulturalHeritageVisit}\\\texttt{Checker.java}}
		& TP &
		\makecell[l]{\texttt{CulturalHeritageVisit}\\\texttt{Checker.java}} \\
\midrule
		
		\texttt{SiteTableModel.java} & FN & --- \\
\midrule
		
		\texttt{ControlData.java} & FN & --- \\
\midrule
		
		--- & FP &
		\makecell[l]{\texttt{TouristClient}\\\texttt{Manager.java}}
		 \\
\midrule
		
		--- & FP &
		\texttt{DBVisitBC.java} \\
\midrule
		
		--- & FP &
		\texttt{BeanVisitBC.java} \\
\midrule
		
		--- & FP &
		\texttt{BeanVisitPR.java} \\
\bottomrule
		
	\end{tabularx}
\caption{Trace links predicted by \tracer (
$\mathcal{P}_1$) for UC$53$.} \label{fig:uc53-tl} 
\vspace{-2em}
\end{figure}

\newtcolorbox[auto counter, number 
within=section]{artifactbox}[2][]{
  colback=gray!10, colframe=black!30,
  title={Example~\thetcbcounter: #2}, fonttitle=\bfseries, #1}

To illustrate, consider the example use case (UC$53$), which  
describes the sequence of actions a tourist performs to submit 
feedback about a particular site. 
In Fig.~\ref{fig:uc53-tl}, we show the trace links associated with 
UC$53$, including both the 
ground-truth links and those predicted by \tracer ($\mathcal{P}_1$). The prediction results contain four TPs, two FNs, 
and four FPs. 
Inspection of the code files associated with the true positives 
suggests that lexical similarity plays a key role in their recovery. 
Specifically, these files are relatively small 
and contain identifiers and Javadoc comments that closely reflect 
the terminology and actions described in UC$53$. 
For example, 
\texttt{IDBVisitBC} and \texttt{IDBVisitPR} are interfaces responsible 
for handling feedback 
and declare the methods \texttt{insertVisitBC} and 
\texttt{insertVisitPR}, respectively, which closely correspond to the 
feedback insertion described in step~6 of the use case.  
Likewise, \texttt{DBVisitPR} implements the insertion of the vote and 
comment provided by the user in step~3, while 
\texttt{CulturalHeritageVisitChecker} 
defines \texttt{checkDataVisitCulturalHeritage}, which performs 
the validation process described in step~4. 
Since these files are relatively small (15 lines of code on 
average) and focused on 
a single responsibility, the domain-specific terms dominate their 
vocabulary, making the trace link recovery easier.

In contrast, the two missed ground-truth links illustrate different 
failure modes. On the one hand, the code file \texttt{SiteTableModel} 
contains no references to concepts such as \emph{feedback}, 
\emph{visit}, \emph{rating}, or \emph{comment}, making it difficult 
to associate UC$53$ with the file through lexical similarity. 
\texttt{ControlData}, on the other hand, is relevant to the 
requirement but devotes only a small portion of its 
implementation (30 out of 281 lines of code) to the visit-validation 
functionality.  
As a result, the traceability signal is present but diluted by unrelated 
functionalities. 

The eTour codebase poses two challenges that affect single-agent 
traceability 
approaches differently. 

First, eTour was originally documented in Italian and later translated 
into English for release as a benchmark. The translation is both 
incomplete and, in some cases, inaccurate. 
Residual Italian text remains in source-code comments
(e.g., Listing~\ref{lst:untranslated} below), while other comments 
contain 
translation errors that alter the intended meaning 
(e.g., Listing~\ref{lst:mistranslation} below).

\begin{lstlisting}[
	style=javaStyle,captionpos=b, 
	caption={Example of a partially untranslated source code 
	comment: ``dell'ArrayList ''},
	label={lst:untranslated},
	xleftmargin=0pt,
	framexleftmargin=0pt
	]
if (statusConvention) {
// Check the contents dell'ArrayList so as not to return
// Null values to the caller
    if (active == null)
        throw new 
        RemoteException(ErrorMessage.ERROR_FORMAT_BEAN);
    return active;
}
\end{lstlisting}

\begin{lstlisting}[
	style=javaStyle, captionpos=b, 
	caption={Example of an incorrectly translated comment: A null 
	value is returned when a match exists rather than when no match 
	is found},
	label={lst:mistranslation},
	xleftmargin=0pt,
	framexleftmargin=0pt
	]
/**
 * Method which returns a banner given its id
 *
 * @Param ID pIdBanner the banner to return
 * @Return Banner found in the database, null if there is' 
 match
 */
public BeanBanner getBannerDaID(int pIdBanner) throws SQLException;
\end{lstlisting}

Overall, \textit{inaccurate documentation has a limited impact on 
retrieval-based approaches} such as LiSSA, as long as the relevant 
terms remain present, because these approaches primarily depend 
on lexical similarity. For example, our analysis shows that 37 of 57 use cases (65\%) share implementation-oriented vocabulary with the identifiers of its linked code files, providing an advantage for retrieval-based approaches. 
In contrast, a single-agent approach 
such as \tracer is \textit{vulnerable to documentation inaccuracies}, 
since it relies on behavioral evidence often conveyed through 
comments. As a result, misleading or incorrect comments may bias 
the model's reasoning process and negatively affect the accuracy of 
its traceability predictions.

Second, the codebase exhibits inconsistent naming conventions. For 
example, \texttt{BeanTourist}, \texttt{DBTourist}, and 
\texttt{TouristAgencyManager} all use role-informative names that 
convey implementation responsibilities (i.e., \texttt{Bean}, 
\texttt{DB}, and \texttt{Manager}). In contrast, \texttt{Tourist.java} 
and \texttt{News.java} are user-interface classes whose names are 
not role-informative, as they refer to domain concepts rather than 
their actual implementation responsibilities.

Again, \textit{retrieval-based approaches like LiSSA are largely 
unaffected by naming inconsistencies} because they rely primarily on 
the broader contents of source files rather than file names. In 
contrast, \tracer uses file names as evidence when identifying 
implementation 
candidates. 
As a result, \textit{the effectiveness of LLM-based traceability 
approaches depends, in part, on the quality and consistency of 
role-informative naming conventions}. 
When names are inconsistent or fail to accurately reflect 
implementation responsibilities, they can bias the model's reasoning 
toward irrelevant artifacts. For instance, a requirement related to 
tourist data may incorrectly suggest \texttt{Tourist.java}, 
despite that the class merely implements a GUI form rather 
than the 
underlying business logic. 

\vspace*{.3em}

\noindent\fbox{\textbf{iTrust}}
Compared to eTour, the iTrust use cases do not follow a consistent 
template. They often omit explicit references to the responsible 
actor and rely on anaphoric expressions that require contextual 
interpretations across requirements. 
As a result, the requirements provide less explicit lexical and 
semantic evidence for 
retrieval-based approaches, such as LiSSA. 
At the same time, iTrust has a well-documented codebase, 
with Javadoc comments
describing the functionality of its classes.
This enables LLM-based approaches such as \tracer to 
compensate for the lack of detail in the
requirements by leveraging information from the source code to infer
the missing context and establish trace links.

To illustrate this challenge, consider the example use case 
UC$6S1$. 
Fig.~\ref{fig:uc6s1-tl} presents the trace links, both the ground-truth 
links and those predicted by \tracer ($\mathcal{P}_1$). 
The results reveal limited accuracy, with only one TP, two FNs, and 
two FPs. 
The only recovered TP, \texttt{ViewVisitedHCPsAction.java}, 
appears to be largely explained by explicit lexical cues. 
The implementation in this file contains concepts directly referenced 
in UC$6S1$, e.g., the HCP name, specialty, 
address, office-visit date, HCP designation, and the generated list of 
visited HCPs. 

\begin{tcolorbox}[ 
	breakable, 
	colback=gray!10, 
	colframe=black!30, 
	left=1mm, right=1mm, 
	top=0.5mm, bottom=0.5mm, 
	boxsep=1mm, 
	arc=1mm,
	fonttitle=\bfseries,
	title={iTrust UC$6S1$},  
	] 
\small
\label{example:iTrust_UC6S1}
The LHCP's name, specialty, address, date of office visit, and whether or not the LHCP is a DLHCP for this patient is indicated. The
list is sorted by the date of the last office visit.
\end{tcolorbox}

\begin{figure}[t]
	\centering
	\footnotesize

	\begin{tabularx}{\linewidth}{
			>{\raggedright\arraybackslash}X
			c
			>{\raggedright\arraybackslash}X}
		\toprule
		\textbf{Ground Truth} & & \textbf{Predicted} \\
		\midrule
		
\makecell[l]{\texttt{ViewVisitedHCPs}\\\texttt{Action.java}} & TP & 
\makecell[l]{\texttt{ViewVisitedHCPs}\\\texttt{Action.java}} \\ \midrule
		
		\texttt{PatientDAO.java} & FN &
		--- \\
		\midrule
		
		\texttt{PersonnelDAO.java} & FN &
		--- \\
		\midrule
		
		--- & FP &
		\texttt{HCPVisitBean.java} \\
		\midrule
		
		--- & FP &
		\texttt{OfficeVisitDAO.java} \\
		\bottomrule
		
	\end{tabularx}
	
	\caption{Trace links predicted by \tracer ($\mathcal{P}_1$) for 
	UC$6S1$}
	\label{fig:uc6s1-tl}
    \vspace{-1.5em}
\end{figure}

On the other hand, the two missed trace links exhibit substantially 
weaker lexical and semantic overlap with the use case. 
Although 
 \texttt{PersonnelDAO.java} is responsible for retrieving the LHCP's 
 name, specialty, and address, 
 the use case does not explicitly identify the entity 
 responsible for providing this information. 
Furthermore, the class refers to 
 ``HCP'' as \emph{Personnel} which cannot be directly inferred from 
 the use case text, and it exposes the relevant attributes (e.g., 
 address) through a bean-loading mechanism. 
Similarly, 
\texttt{PatientDAO.java} implements the DLHCP check through 
using the concept of a \emph{DeclaredHCP},  a term that is only 
implicitly related to the use case through the anaphoric reference 
\emph{this patient}.
This semantic gap weakens the connection between the use case 
and the source code.  
In addition, \texttt{PatientDAO.java}  contains 765 lines, 
spanning multiple unrelated 
responsibilities, which further dilutes the relevant traceability 
evidence.

The vocabulary needed to identify both code files is present in other 
use cases, namely UC$6S2$ and UC$6S3$ shown below. 
UC$6S2$, for instance, describes 
the designation and undesignation of an LHCP as a DLHCP, 
corresponding 
to the methods \texttt{declareHCP} and \texttt{undeclareHCP} 
implemented in
\texttt{PatientDAO.java}. 
UC$6S3$ describes searching by last name and zip code, 
functionality implemented by 
\texttt{searchForPersonnelWithName} in 
\texttt{PersonnelDAO.java}.  
Although these three use cases describe complementary  aspects of 
the same functionality, they are traced independently.
Consequently, the evidence supporting the correct links is dispersed 
across 
multiple requirements in the dataset rather than being concentrated 
in a single use case. This 
illustrates the challenge of \emph{scattered requirements}, where 
information related to a feature is 
fragmented across several requirements. Following this pattern, we identified 97 scattered use cases out of the total 105 use cases considered in our analysis (92\%).
This weakens the 
lexical and semantic relatedness between individual requirements 
and the code files that implement the feature~\cite{tobias2026}.

\begin{tcolorbox}[ 
	breakable, 
	colback=gray!10, 
	colframe=black!30, 
	left=1mm, right=1mm, 
	top=0.5mm, bottom=0.5mm, 
	boxsep=1mm, 
	arc=1mm,
	fonttitle=\bfseries,
	title={iTrust UC$6S2$},  
	] 
	\small
\label{example:iTrust_UC6S2}
The patient can choose to toggle between 
designating/undesignating any LHCP as being a DLHCP for them.
\end{tcolorbox}

\begin{tcolorbox}[ 
	breakable, 
	colback=gray!10, 
	colframe=black!30, 
	left=1mm, right=1mm, 
	top=0.5mm, bottom=0.5mm, 
	boxsep=1mm, 
	arc=1mm,
	fonttitle=\bfseries,
	title={iTrust UC$6S3$},  
	] 
	\small
\label{example:iTrust_UC6S3}
The patient types a last name or partial last name, and/or providing 
the specialty. The patient may optionally enter a zip code (match
on first three numbers of zip code), in addition to the name and/or 
specialty. The LHCP's name, specialty, and address are provided.
\end{tcolorbox}

The iTrust codebase exhibits characteristics largely opposite to 
those of eTour. 
It is written entirely in English, and 216 of its 226 classes contain 
Javadoc comments that describe what a
method does, what its parameters mean, and what it returns. 
Furthermore, informative naming conventions are applied 
consistently throughout the codebase, with suffixes such as 
\emph{Action},
\emph{DAO}, \emph{Bean}, \emph{Loader}, and \emph{Validator}, 
clearly conveying architectural roles. 

These characteristics benefit single-agent approaches 
like \tracer, which reason directly over the source code. 
Generally, \textit{accurate documentation and consistent naming 
conventions provide stronger traceability evidence. }
These cues reduce ambiguity, narrow the search space, and support 
more effective 
reasoning about how requirements are implemented in the code.
Retrieval-based approaches like LiSSA, however, benefit far less  
from these code quality aspects because its primary limitation lies in 
the 
requirements themselves rather than the implementation artifacts. 
Consequently, \textit{the effectiveness of retrieval-based 
approaches is constrained by the amount of explicit lexical evidence 
available in the requirements}. 
When relevant information is missing, implicit, or distributed across 
multiple requirements like is the case with iTrust use cases, even 
well-documented and consistently named code files provide limited 
additional benefit. This is reflected in 
LiSSA (IR-Only), which achieves 
53.1\% recall.

\vspace*{.3em}
\noindent\fbox{\textbf{Dronology}}
Dronology sits between eTour and iTrust in terms of requirements 
quality. 
Most requirements follow a concise ``shall''-style format that 
explicitly specifies 
the responsible component and describes a single behavior, 
providing clear lexical and semantic cues for traceability.  

To illustrate, consider requirement RE$28$, together with its 
ground-truth and predicted trace links presented in 
Fig.~\ref{fig:re28-tl}.  

\begin{tcolorbox}[ 
	breakable, 
	colback=gray!10, 
	colframe=black!30, 
	left=1mm, right=1mm, 
	top=0.5mm, bottom=0.5mm, 
	boxsep=1mm, 
	arc=1mm,
	fonttitle=\bfseries,
	title={Dronology RE$28$}, 
	] 
	\small 
\textbf{[SUMMARY]} UAV Flight Plan Execution\par
\textbf{[DESCRIPTION]} The \emph{SingleUAVFlightPlanScheduler} shall only execute
one flight plan at a time for each UAV.
\end{tcolorbox}

\begin{figure}[t]
	\centering
	\footnotesize

	\begin{tabularx}{\linewidth}{
			>{\raggedright\arraybackslash}X c
			>{\raggedright\arraybackslash}X}
		\toprule
		\textbf{Ground Truth} & & \textbf{Predicted} \\
		\midrule
		
		\texttt{PlanPool.java} & TP &
		\texttt{PlanPool.java} \\
		\midrule
		
		\texttt{PlanPoolManager.java} & TP &
		\texttt{PlanPoolManager.java} \\
		\midrule
		
		--- & FP &
		\texttt{FlightZoneManager2.java} \\
		\midrule
		
		--- & FP &
		\texttt{FlightInfo.java} \\
		\bottomrule
		
	\end{tabularx}
	
	\caption{Trace links predicted by \tracer{} ($\mathcal{P}_1$) for 
	RE$28$.}
	\label{fig:re28-tl}
	\vspace{-1.5em}
\end{figure}

Here, once again, we consider the predictions 
generated by \tracer ($\mathcal{P}_1$). The results contain two TPs, 
no FNs, and two FPs.
Both ground-truth links are recovered primarily through strong 
lexical and semantic alignment between the requirement and the 
implementation artifacts.
Specifically, the requirement provides three traceability cues: 
(i) the component name \texttt{SingleUAVFlightPlanScheduler}, for 
which no corresponding class
exists in the codebase; 
(ii) the constraint that \emph{only one flight plan may be executed at 
a time}; and (iii) 
the quantifier \emph{each UAV}. 

Inspection of the implementation reveals strong correspondence 
between these cues and the associated code files. 
The \texttt{PlanPool} class is 
documented as managing
``pending, \emph{executing}, and completed plans for a 
\emph{single UAV}'',  
and contains the only occurrence of the phrase \emph{single UAV} 
in the codebase.  
Furthermore, 
it maintains a single \texttt{currentPlan} and provides methods 
such as \texttt{addPlan} and \texttt{setNextPlanToCurrent}, directly 
reflecting the requirement's scheduling constraint. 
Similarly, \texttt{PlanPoolManager} maintains a \texttt{PlanPool} for 
``each active UAV'' and organizes plans by UAV identifier. 
The requirement is therefore largely restated in the implementation 
vocabulary, making retrieval more straightforward without minimal 
reasoning over the code files. This example illustrates how explicit 
correspondence between requirement terminology and 
implementation concepts can substantially simplify traceability link 
recovery.

A small subset of the Dronology requirements (4 out of 56) omits the 
responsible component and specifies only a system-wide data 
constraint. 
As a result, these 
requirements are implemented across many classes, producing 
an unusually large number of trace links. 
This suggests that \textit{cross-cutting requirements are intrinsically 
more challenging to trace than requirements describing localized 
behavior}, since the relevant implementation evidence is scattered 
across multiple artifacts. Consequently, these requirements can have 
a disproportionate impact on aggregate performance metrics.

\begin{tcolorbox}[ 
	breakable, 
	colback=gray!10, 
	colframe=black!30, 
	left=1mm, right=1mm, 
	top=0.5mm, bottom=0.5mm, 
	boxsep=1mm, 
	arc=1mm,
	fonttitle=\bfseries,
	title={Dronology RE$77$}, 
	] 
	\small 
\textbf{[SUMMARY]} Waypoint definition\par
\textbf{[DESCRIPTION]} Each waypoint shall be specified using coordinates that include longitude  latitude  and altitude.
\end{tcolorbox}

\begin{figure}[t]
	\centering
	\footnotesize

	\begin{tabularx}{\linewidth}{
			>{\raggedright\arraybackslash}X c
			>{\raggedright\arraybackslash}X}
		\toprule
		\textbf{Ground Truth} & & \textbf{Predicted} \\
		\midrule
		
		\texttt{Waypoint.java} & TP &
		\texttt{Waypoint.java} \\
		\midrule
		
		\texttt{LlaCoordinate.java} & TP &
		\texttt{LlaCoordinate.java} \\
		\midrule
		
		\makecell[l]{
			\texttt{AbstractPosition.java},\\
			\texttt{FlightRoute.java},\\
			\texttt{FlightPlan.java},\\
			\texttt{ManagedDrone.java},\\
			\texttt{NvecInterpolator.java},\\
			\texttt{GoToCommand.java},\\
			\texttt{DronologyConstants.java},\\
			\texttt{DummyLockObject.java},\\
			\ldots{} and 64 additional classes
		}
		& FN &
		--- \\
		\bottomrule
		
	\end{tabularx}
	
	\caption{Trace links predicted by \tracer{} ($\mathcal{P}_1$) for 
	RE$77$.}
	\label{fig:re77-tl}
	\vspace{-1.5 em}
\end{figure}

To illustrate this challenge, consider the example requirement RE$77$ 
and 
the corresponding trace links in 
Fig.~\ref{fig:re77-tl}. 

Unlike RE$28$, RE$77$ does not specify a responsible component; 
instead, it defines only a system-wide data constraint. 
Although 
relevant vocabulary is still present in the codebase, it is concentrated 
in only a small subset of the implementation artifacts. For example,  
\texttt{LlaCoordinate} is 
documented as ``a terrestrial position 
defined by \emph{latitude}, \emph{longitude}, and \emph{altitude}'' and declares 
exactly those three fields. Likewise, \texttt{Waypoint} is documented 
as representing coordinates using  \texttt{LlaCoordinate}. 
These explicit correspondences enable \tracer to correctly recover 
both code files as valid trace links.
However, the remaining ground-truth links are associated with classes that realize the 
constraint only indirectly. As a result, requirement RE$77$ 
provides limited evidence for identifying the full implementation 
scope, yielding only two TPs and 74 FNs out of 184 code files. The 
presence of four similarly underspecified requirements in Dronology 
helps explain the lower overall performance observed on this dataset. 

This example highlights that \textit{requirements specifying  
generic 
data constraints often provide only weak evidence for the 
numerous artifacts 
that collectively implement those constraints}.

\begin{figure}[t]
	\centering
	\footnotesize
	
	\begin{tabularx}{\linewidth}{
			>{\raggedright\arraybackslash}X c
			>{\raggedright\arraybackslash}X}
		\toprule
		\textbf{Ground Truth} & & \textbf{Predicted} \\
		\midrule
		
		\texttt{Waypoint.java} & TP &
		\texttt{Waypoint.java} \\
		\midrule
		
		\texttt{LlaCoordinate.java} & TP &
		\texttt{LlaCoordinate.java} \\
		\midrule
		
		\makecell[l]{
			\texttt{AbstractPosition.java},\\
			\texttt{FlightRoute.java},\\
			\texttt{FlightPlan.java},\\
			\texttt{ManagedDrone.java},\\
			\texttt{VirtualDrone.java},\\
			\texttt{GoToCommand.java},\\
			\ldots{} and 20 additional classes
		}
		& TP &
		\makecell[l]{
			\texttt{AbstractPosition.java},\\
			\texttt{NvecInterpolator.java},\\
			\texttt{FlightRoute.java},\\
			\texttt{FlightRouteInfo.java},\\
			\texttt{IFlightRoute.java},\\
			\texttt{FlightPlan.java},\\
			\texttt{FlightPlanFactory.java},\\
			\texttt{FlightPlanInfo.java},\\
			\texttt{IFlightPlan.java},\\
			\texttt{FlightInfoCreator.java},\\
			\texttt{SoloDirector.java},\\
			\texttt{ManagedDrone.java},\\
			\texttt{VirtualDrone.java},\\
			\texttt{PhysicalDrone.java},\\
			\texttt{GoToCommand.java},\\
			\ldots{} and 11 additional classes
		}
		\\
		\midrule
		
		\makecell[l]{
			\texttt{DronologyConstants.java},\\
			\texttt{DummyLockObject.java},\\
			\ldots{} and 46 additional classes
		}
		& FN &
		--- \\
		\midrule
		
		--- & FP &
		\makecell[l]{
			\texttt{FlightRouteInfo.java},\\
			\texttt{IFlightRoute.java},\\
			\texttt{FlightPlanFactory.java},\\
			\texttt{FlightPlanInfo.java},\\
			\texttt{IFlightPlan.java},\\
			\texttt{FlightInfoCreator.java},\\
			\texttt{SoloDirector.java},\\
			\texttt{PhysicalDrone.java},\\
			\texttt{FlightZoneManager2.java},\\
			\makecell[l]{\texttt{DronologyPersistence}\\\texttt{Util.java}}\\
			\ldots{} and 16 additional classes
		}
		\\
		\bottomrule
		
	\end{tabularx}
	
	\caption{Trace links predicted by \tracer{} ($\mathcal{P}_2$) for 
	RE$77$.}
	\label{fig:re77-tl-p2}
	\vspace{-1.5em}
\end{figure}

While $\mathcal{P}_1$ is conservative, a more recall-oriented prompt 
like $\mathcal{P}_2$ is particularly beneficial in these cases. 
As shown in Fig.~\ref{fig:re77-tl-p2}, using a more recall-oriented 
prompt ($\mathcal{P}_2$) recovers many additional links for RE$77$, 
including 
classes whose relevance is evident only from their implementation 
rather than by their
file naming or documentation. 

For example, \texttt{GoToCommand.java} never references a 
\textit{waypoint} in its documentation, yet its constructor 
decomposes a coordinate into the required 
\texttt{x}, \texttt{y}, and \texttt{z} components. Similarly,
\texttt{VirtualDrone.java} does not mention \textit{waypoints}, but 
it manages UAV positions through methods such as  
\texttt{setCoordinates(latitude, longitude, altitude)}. Recovering 
these links therefore requires reasoning about the functionality 
implemented by the code rather than 
relying on explicit terminology in class names or documentation. 
The more permissive $\mathcal{P}_2$ prompt encourages such 
reasoning, enabling the recovery of links supported only by implicit 
evidence, whereas $\mathcal{P}_1$ tends to discard such candidates 
because the requirement provides only limited lexical and semantic 
cues. The difference in recall supports this finding: R=25.7\% for 
$\mathcal{P}_1$ compared to R=69.7\% for $\mathcal{P}_2$. 

The Dronology codebase shares more characteristics with iTrust 
rather than eTour. 
It is written entirely in English and employs consistent naming 
conventions: 
30 interfaces carry the prefix \emph{I}, and suffixes such as
\emph{Service}, \emph{Manager} and \emph{Command} clearly 
indicate  the architectural responsibilities, 
with 120 of its 184 classes adhering to these role-informative naming 
conventions. 
Although documentation is less comprehensive than in iTrust, with 
only 92 (out of 184) classes containing class-level comments, this 
limitation is 
partially mitigated by the descriptive nature of the identifiers used 
throughout the codebase. As a result, \textit{behavioral and 
architectural information is often conveyed directly through class, 
method, and field names rather than through documentation}.
Even 
undocumented classes, such as \texttt{FlightInfo}, remain largely 
self-descriptive through names such as \texttt{currentFlight}, 
\texttt{pending}, \texttt{completed}, and \texttt{addPending}, which 
provide strong cues about their responsibilities and behavior. 

Dronology is therefore the only dataset in which the code artifacts 
consistently provide high-quality traceability evidence, while the 
requirements are comparatively less informative.  In 
particular, requirements often refer to conceptual components, such 
as \texttt{SingleUAVFlightPlanScheduler}, that have no 
direct implementation artifact. 
As a result, \textit{the terminology used in the requirements does not 
always align with the vocabulary of the codebase}, weakening the 
lexical correspondence on which retrieval-based approaches depend. 
This challenge is reflected in LiSSA's performance, with the IR-only 
variant achieving a recall of only 44.8\%.

In Dronology, after excluding underspecified requirements, 53 requirements (93\%) can be traced because of the explicit naming information provided in the requirements, while the remaining four requirements (7\%) are underspecified and therefore associated with a large number of trace links.

\subsubsection*{\textbf{Key takeaway}} 
The qualitative analysis suggests that the effectiveness of traceability 
approaches depends largely on where the most informative 
traceability evidence resides. 

\textit{Retrieval-based approaches perform best when requirements 
contain explicit implementation-oriented terminology that closely 
aligns with the vocabulary of the codebase.} This explains LiSSA's 
strong performance on eTour, where requirements exhibit substantial 
lexical overlap with the corresponding implementation artifacts. 

In contrast, \textit{single-agent approaches that reason 
over the entire codebase benefit most from high-quality 
implementation artifacts, including accurate documentation, 
consistent role-informative naming conventions, and descriptive 
identifiers.} This explains \tracer's superior performance on iTrust and 
Dronology, where the codebase provides rich behavioral and 
architectural evidence. By reasoning directly over source code, \tracer 
can exploit information that is only implicitly related to the requirement 
and therefore remains inaccessible to purely retrieval-based 
methods. 

Overall, the results indicate that \textit{retrieval-based approaches are 
primarily constrained by requirement quality, whereas code-centric 
LLM-based approaches are primarily influenced by code quality}. 
Consequently, \tracer is more resilient to requirement quality issues 
such as anaphoric references, omitted actors, scattered 
requirements, and conceptual components that lack direct 
implementation counterparts, provided that the corresponding 
evidence is available in the codebase.

\subsection{Qualitative Comparison of Single- and 
Multi-Agent Approaches}
The choice between single-agent and multi-agent approaches 
ultimately depends on the analyst's validation budget and 
available computational resources. 
In the following, we qualitatively analyzed a subset of 55 use cases out of the total 105 from iTrust (42\%) to better understand 
the impact of multi-agent reasoning by analyzing the output of 
\tracer ($\mathcal{P}_1$) compared to \debater 
($\mathcal{P}_1$), as well as the contribution of the selected  
prompt by comparing \debater ($\mathcal{P}_1$) against  
\debater 
($\mathcal{P}_2$).
To illustrate this process, we consider use case UC$26S2$ 
from iTrust, shown below along with its six associated 
ground-truth code files.
\begin{tcolorbox}[ 
	breakable, 
	width=\linewidth,
	colback=gray!10, 
	colframe=black!30, 
	left=1mm, 
	right=1mm, 
	top=0.5mm, 
	bottom=0.5mm, 
	boxsep=1mm, 
	arc=1mm,
	fonttitle=\bfseries,
	title={iTrust UC$26S2$},  
	] 
	\small
	\label{example:iTrust_UC26S2}
	
	An HCP can view a previously created lab procedure for a 
	given office visit. The HCP can view patient name, lab 
	procedure code,
	current lab procedure status, timestamp, and Lab Technician 
	name.

\end{tcolorbox}

Fig.~\ref{fig:UC26S2-link-tracer} shows the trace links 
recovered by \tracer ($\mathcal{P}_1$). \tracer 
recovers only one correct link for UC$26S2$, 
namely \texttt{LabProcedureDAO.java}. 
This is also the only code file whose documentation restates 
the use case text.
Consequently, establishing this link requires limited reasoning  
beyond straightforward lexical matching between the use case 
and the code documentation.

The five missed code files do not contain documentation that 
explicitly uses the same vocabulary as the use case.
For instance, the attributes \emph{patient name} and 
\emph{Lab Technician name} are not stored by the lab 
procedure itself, which retains only the numeric identifiers 
\texttt{pid} and \texttt{ovID}; the corresponding names must 
therefore be obtained through a separate lookup.
This functionality is implemented in 
\texttt{GetUserNameAction.java} and \texttt{AuthDAO.java}, 
whose documentation refers only to a \textit{person's name 
}or a \textit{user name} and never mentions lab procedures, 
office visits, or HCPs. 
Identifying these code files as candidate links requires 
reasoning about how the displayed attributes are retrieved 
rather than matching the terminology used in the use case.  
Moreover, this relationship can only be inferred by combining 
information across multiple code files.

A similar issue arises with 
\texttt{LabProcUAPAction.java}. Although its documentation 
mentions \textit{lab procedure}, 
it is written for \textit{UAPs} rather than the \textit{HCPs} 
referenced in the use case. Thus, the file appears both 
relevant and irrelevant at the same time. \tracer cannot 
reconcile these conflicting cues in a single-pass decision. 

\begin{figure}[tp]
	\centering
	\footnotesize
	
	\begin{tabularx}{\linewidth}{
			>{\raggedright\arraybackslash}X
			c
			>{\raggedright\arraybackslash}X}
		\toprule
		\textbf{Ground Truth} & & \textbf{Predicted} \\
		\midrule
		
		\texttt{LabProcedureDAO} & 
TP 
		&
		\texttt{LabProcedureDAO}  \\
		\midrule
		\texttt{GetUserNameAction} & 
		FN &
		--- \\
		\midrule
		
		\texttt{AuthDAO.java} & FN &
		--- \\
		\midrule
		
	\texttt{LabProcUAPAction} & 
FN 
		&
		--- \\
		\midrule
		
		\texttt{EditPatientAction} & FN 
&
		--- \\
		\midrule
		
		\texttt{PatientDAO} & FN &
		--- \\
		\midrule
		
		--- & FP &
		\texttt{EditOfficeVisitAction} \\
		\midrule
		
		--- & FP &
		\texttt{LabProcedureBean} \\
		\midrule
		
		--- & FP &
		
  		\texttt{LabProcedureBeanLoader}
		 \\
		\midrule
		
		--- & FP &
		\texttt{OfficeVisitDAO} \\
		\bottomrule
		
	\end{tabularx}
	
	\caption{\tracer ($\mathcal{P}_1$) predicted files 
		(\texttt{.java}) 
		for  UC$26S2$. }
	\label{fig:UC26S2-link-tracer}
    \vspace{-1.5em}
\end{figure}

Fig.~\ref{fig:UC26S2-link-debater-p1} presents the 
trace links predicted by \debater ($\mathcal{P}_1$). In addition 
to recovering 
\texttt{LabProcedureDAO.java} (identified by \tracer), \debater 
correctly recovers two 
more ground-truth code files: 
\texttt{LabProcUAPAction.java} and 
\texttt{AuthDAO.java}. 
Neither file restates the use case as directly as 
\texttt{LabProcedureDAO.java}, and each is missed by \tracer 
for a different reason as explained earlier. In both cases, 
however, the necessary 
evidence is present in the code and documentation, albeit in a 
weaker and less explicit form.

\begin{figure}[t]
	\centering
	\footnotesize
	
	\begin{tabularx}{\linewidth}{
			>{\raggedright\arraybackslash}X
			c
			>{\raggedright\arraybackslash}X}
		\toprule
		\textbf{Ground Truth} & & \textbf{Predicted} \\
		\midrule
		
		\texttt{LabProcedureDAO}& 
TP 
		&
		\texttt{LabProcedureDAO} \\
		\midrule
		
		\texttt{LabProcUAPAction} & 
TP 
		&
		\texttt{LabProcUAPAction} \\
		\midrule
		
		\texttt{AuthDAO} & TP &
		\texttt{AuthDAO} \\
		\midrule
		
		\texttt{GetUserNameAction} & 
		FN &
		--- \\
		\midrule
		
		\texttt{PatientDAO} & FN &
		--- \\
		\midrule
		
		\texttt{EditPatientAction} & FN 
&
		--- \\
		\midrule
		
		--- & FP &
		\texttt{EditOfficeVisitAction} \\
		\midrule
		
		--- & FP &
		\texttt{LabProcedureBean}\\
		\midrule
		
		--- & FP &
		
		\texttt{LabProcedureBeanLoader}
		 \\
		\midrule
		
		--- & FP &
		\texttt{OfficeVisitBaseAction}\\
		\midrule
		
		--- & FP &
		\texttt{PersonnelDAO} \\
		\bottomrule
		
	\end{tabularx}
	
	\caption{\debater ($\mathcal{P}_1$) predicted files 
				(\texttt{.java}) 
				for  UC$26S2$. }
	\label{fig:UC26S2-link-debater-p1}
    \vspace{-1.5em}
\end{figure}

\texttt{LabProcUAPAction.java} shares several key concepts 
with the use case. It declares a \texttt{LabProcedureDAO}  
field, and its \texttt{viewProcedures(id)}  method is 
documented as returning ``a list of all the lab procedures for 
that UAP." The main source of ambiguity is the role qualifier 
\textit{UAP}, which differs from the \textit{HCP} referenced in 
the use case. As a result, the file contains both supporting and 
conflicting signals. Unlike \tracer, \debater revisits the 
candidate after establishing the supporting evidence, allowing 
the role conflict to be interpreted in the broader 
implementation context.

\texttt{AuthDAO.java} presents the opposite challenge. Its 
documentation contains no vocabulary related to lab 
procedures and instead describes the class as being ``for 
anything that has to do with authentication'', making it more 
unlikely to be selected by a single-pass tracing approach. Its 
relevance stems from a single method 
\texttt{getUserName(mid)}, which is documented as 
returning ``the user name of a user from just the MID''. 
This method becomes important when examining how the 
information required by the use case is obtained. 
Two of the five attributes listed in the use case, \textit{patient 
name} and \textit{Lab Technician name}, are not stored 
directly in the lab procedure record. 
Instead, the record stores only numeric identifiers, while the 
remaining attributes, namely the \textit{procedure code}, 
\textit{status}, and \textit{timestamp}, are available directly.
Consequently, displaying the two names requires resolving 
identifiers into user names, and \texttt{getUserName(mid)} 
provides exactly this functionality. Although the connection is 
indirect and not expressed through shared terminology, the 
method is necessary for implementing the behavior described 
by the use case.

\begin{figure}[t]
	\centering
	\footnotesize
	
	\begin{tabularx}{\linewidth}{
			>{\raggedright\arraybackslash}X
			c
			>{\raggedright\arraybackslash}X}
		\toprule
		\textbf{Ground Truth} & & \textbf{Predicted} \\
		\midrule
		
		\texttt{LabProcedureDAO} & 
TP 
		&
		\texttt{LabProcedureDAO} \\
		\midrule
		
		\texttt{LabProcUAPAction}& 
TP 
		&
\texttt{LabProcUAPAction}\\
		\midrule
		
		\texttt{AuthDAO} & TP &
		\texttt{AuthDAO} \\
		\midrule
		
		\texttt{PatientDAO} & TP &
		\texttt{PatientDAO} \\
		\midrule
		
		\texttt{GetUserNameAction} & 
		FN &
		--- \\
		\midrule
		
 	\texttt{EditPatientAction} & FN 
&
		--- \\
		\midrule
		
		--- & FP &
 \texttt{EditOfficeVisitAction}\\
		\midrule
		
		--- & FP &
\texttt{LabProcedureBean}\\
		\midrule
		
		--- & FP &
		
\texttt{LabProcedureBeanLoader}		 \\
		\midrule
		
		--- & FP &
		\texttt{LOINCDAO} \\
		\midrule
		
		--- & FP &
		\texttt{LOINCbean} \\
		\midrule
		
		--- & FP &
		\texttt{LOINCBeanLoader} \\
		\midrule
		
		--- & FP &
		\texttt{OfficeVisitBaseAction} \\
		\midrule
		
		--- & FP &
		\texttt{OfficeVisitDAO} \\
		\midrule
		
		--- & FP &
		\texttt{OfficeVisitBean} \\
		\midrule
		
		--- & FP &
		\texttt{PersonnelDAO} \\
		\midrule
		
		--- & FP &
		\texttt{DAOFactory} \\
		\midrule
		
		--- & FP &
 		\texttt{TransactionType} \\
		\bottomrule
		
	\end{tabularx}
	
	\caption{\debater ($\mathcal{P}_2$) predicted files 
	(\texttt{.java}) 
	for  UC$26S2$. }
\label{fig:UC26S2-link-debater-p2}
\vspace{-1.5em}
\end{figure}

\debater ($\mathcal{P}_2$) applies the same 
debate-and-correction strategy as \debater with 
$\mathcal{P}_1$ and recovers the same three links, \texttt{LabProcedureDAO.java}, 
\texttt{LabProcUAPAction.java} and \texttt{AuthDAO.java}, for 
essentially the same reasons. 
In addition, it identifies \texttt{PatientDAO.java}. This file 
exhibits no lexical 
overlap with the use case: its documentation states that it 
manages 
``all static information related to a patient'' and contains no 
references to lab procedures, office visits 
or HCPs. Thus, neither its documentation nor its comments 
suggest the behavior required by its use 
case. 
However, the class provides  the method 
\texttt{getName(mid)}, which is documented to return 
``the patient's first name and last name'' for a given MID. This 
method performs the actual name lookup that 
\texttt{AuthDAO.getUserName(mid)} delegates 
to after determining that the identifier belongs to a patient. 

The additional link is recovered because the more  
recall-oriented prompt $\mathcal{P}_2$ admits candidates 
whose relationship to 
the use case is indirect and observable only through  the call 
dependencies rather than lexical evidence.
Although this broader search space enables the discovery of 
\texttt{PatientDAO.java}, it also increases the number of files 
predicted for UC$26S2$ from 
eight to sixteen.

Out of the 55 iTrust use cases analyzed, \debater~($\mathcal{P}_1$) recovered ground-truth links missed by \tracer~($\mathcal{P}_1$) in 12 cases (21.8\%), while 
\debater~($\mathcal{P}_2$) recovered additional links missed by \debater~($\mathcal{P}_1$) in eight cases (14.5\%).

\subsubsection*{\textbf{Key takeaway}} 

In the cases we examined, the three configurations differ in the kind of code evidence they can exploit. \textit{Single-agents with a conservative prompt, like \tracer ($\mathcal{P}_1$),
reliably recover links when the relevance of a code file is explicit}, such as through documentation or identifiers that closely match the use case. However, \textit{they often miss files whose relevance must be inferred}, e.g., when the supporting cues are contradictory or confined to a single method within an otherwise unrelated class. \textit{Using the same prompt, multi-agent \debater ($\mathcal{P}_1$) recovers some of these links,} because the \textit{trace debater agent} revisits files rejected in the initial round, allowing weak or conflicting evidence to be reconsidered before the \textit{judge agent} decides. \textit{With the recall-oriented prompt ($\mathcal{P}_2$), \debater additionally recovers links whose relevance is not documented but emerges from the implementation itself,} e.g., through method-level dependencies and delegated calls that resolve identifiers across classes. \textit{Relaxing the prompt can partially enable this behavior in a single agent setting}, as observed in the case of  Dronology, where \tracer ($\mathcal{P}_2$) recovered code files 
that do not mention the requirement's terminology. 
\textit{Each step toward weaker evidence, however, admits more false positives.} For example, moving from \debater ($\mathcal{P}_1$) to \debater ($\mathcal{P}_2$) recovers one additional true link for UC26S2 in iTrust, but also doubles the number of predicted files, from eight to sixteen. \textit{Some links} 
\textit{remain unrecovered by all configurations, indicating cases where the available evidence is insufficient even under more permissive reasoning. }

\renewcommand{\thefootnote}{\arabic{footnote}}

\section{Related Work} 
\label{sec:related}

Traditionally, TLR has been studied through two main paradigms: 
(i)~\emph{information retrieval}, which establishes trace links 
using semantic similarity metrics and lexical overlap~\cite{antoniol2002,cleland2003,marcus2003,borg2014}; and 
(ii)~\emph{machine learning}, where supervised models are trained 
on traceability datasets to predict trace links~\cite{hayes2006,gethers2011,shaukat2018,mills2018,guo2017,Rasiman2022}. 
However, these approaches often struggle to capture rich 
relationships among heterogeneous software artifacts and 
typically rely on costly, manually labeled data. Recent advances 
in LLMs have reshaped TLR research by enabling more effective 
reasoning across heterogeneous software artifacts (e.g., 
natural-language requirements and source code). Beyond single-agent prompting, multi-agent approaches, in which several LLM-based agents with distinct roles collaborate on a task, have emerged as a promising way to address complex software engineering problems. 
In the 
following, we position our work with respect to emerging 
LLM-supported requirement-to-code traceability.

\noindent \emph{\textbf{LLMs for Traceability.}} 
With the exception of a few studies on supervised fine-tuning of 
LLMs (e.g.,~\cite{ge2025}), most existing research relies on 
prompt engineering and retrieval-augmented generation (RAG). 
Prior work investigates both manual prompt engineering~\cite{rodriguez2023,hassine2024} 
and automated prompt optimization~\cite{north2024,alturayeif2026} 
for identifying trace links across software artifacts. 
Complementary studies leverage established prompting 
frameworks and templates to structure traceability 
analysis~\cite{vogelsang2025,etezadi2026}. The literature also 
emphasizes the sensitivity of prompt-based approaches: small 
wording changes can substantially alter the produced trace 
links, making performance depend not only on model 
capabilities but also on prompt quality, whose impact is difficult to assess without extensive empirical evaluation.~\cite{rodriguez2023,alturayeif2026}.

Because prompts must often include the artifacts under analysis 
along with task instructions, they can consume a large fraction 
of the available context window\footnote{While recent LLMs support substantially larger context windows, this constraint was a significant limitation in earlier models and can still pose challenges for large software artifacts.}. To mitigate this limitation, 
existing work adopts RAG, which first retrieves a small set of 
relevant target artifacts and then applies LLM-based reasoning 
only to those candidates~\cite{wang2025}. For instance, Huang 
et al.~\cite{huang2026} retrieve and rerank candidate commits 
for issue-to-commit traceability; Ali et al.~\cite{ali2024} combine 
retrieval, dependency information, and LLM reasoning for 
requirements-to-code traceability; and Fuch\ss{} et al.~\cite{fuchbeta2025} 
propose LiSSA, a generic RAG-based approach applicable 
across artifact types (discussed in Section~\ref{subsec:baseline}). 
In a follow-up study on inter-requirement TLR, Fuch\ss{} et al.~\cite{fuchss2025beyond} 
replace LiSSA's retrieval step with an ensemble of small LLMs 
that incrementally filters out unrelated requirement pairs 
before passing the remaining candidates to an LLM classifier 
for final trace-link prediction.

Despite their differences, these approaches share a key limitation: they judge trace links \emph{pairwise}. In the requirement-to-code traceability setting considered here, each code file is assessed largely in isolation from the rest of the codebase. This is evident in explicit classification pipelines, where an LLM predicts one requirement--code pair at a time~\cite{fuchss2025beyond}. The same limitation applies to RAG-based methods, where retrieval ranks each code file by its individual similarity to the requirement~\cite{ali2024,huang2026}.
Yet requirements are rarely realized by a single code file. Instead, 
their behavior emerges from collaborating classes, many of which resemble the requirement only weakly when inspected in isolation. Contextual information is required to better understand the contribution of a file to the overall implementation to identify trace links more accurately. Such contextual information is exactly what pairwise approaches discard.

Our work takes a different approach: we provide the LLM with both the requirement and the full codebase. Rather than reasoning over isolated files or a small set of retrieved candidates, the model can consider the repository as a whole and leverage structural and contextual relationships among files. 
To attribute any improvement to full-code context rather than prompt design, 
we use two zero-shot prompts from prior work and apply 
both to our approach and to the retrieval-based baseline, with the same 
model. The first is the chain-of-thought prompt of the 
baseline~\cite{fuchbeta2025}. The second adds a single instruction to it: to 
prioritize recall over precision, following Etezadi 
et al.~\cite{etezadi2026}, who included this instruction in their 
engineered traceability prompt because rejecting false positives costs 
analysts less effort than finding missed links. Our approach and the 
baseline thus differ only in their input (Section~\ref{sec:eval}).

\noindent \emph{\textbf{LLM-based Multi-agent Approaches in Software 
Engineering. }}
Multi-agent systems have a long history in software engineering research. Early work, e.g.,~\cite{wooldridge2000,deloach2001,bresciani2004} treated agents as the software systems being engineered. In contrast, recent LLM-based multi-agent approaches use agents as a mechanism for performing software engineering tasks. Building on this foundation, LLM-based multi-agent approaches have been applied to various tasks, including code generation~\cite{huang2024,ishibashi2024,hong2024}, code review~\cite{tang2024}, and issue resolution~\cite{chen2024teaching,tao2024}. 

The use of multi-agents for traceability remains largely underexplored. A notable exception is TraceDev by Chen et al.~\cite{chen2026}, which integrates traceability into multi-agent code generation. TraceDev generates repository-level code from use cases using five agents: a requirement refiner; a designer producing Unified Modeling Language (UML) class and sequence diagrams; a developer; a tester; and a validator. The validator links requirements to design elements via an LLM, and links design elements to code by checking whether the specified classes and methods are defined or invoked in the generated code. Missing links are fed back to the designer and developer for refinement. Although TraceDev has been evaluated on the eTour and SMOS traceability datasets, its goal is to generate new code rather than recover trace links in an existing codebase; accordingly, it assesses generated-code quality rather than trace-link accuracy.

Our work examines two multi-agent approaches for trace-link recovery. The first builds on self-correction, in which an LLM iteratively refines its output based on feedback~\cite{madaan2023,chen2024teaching}. In particular, we draw on Reflexion by Shinn et al.~\cite{shinn2023}, where an agent reflects on feedback from failed attempts and uses these reflections to guide subsequent attempts.

The closest related work is R2Code by Wang et al.~\cite{wang2026}, which adds self-verification to requirements-to-code trace-link recovery. An LLM first summarizes each code unit, then retrieves candidates for each requirement whose term overlap with the summary exceeds a threshold. Next, it decomposes the requirement and each candidate into semantic layers and scores their alignment pairwise. Finally, it explains each predicted link, and a second LLM call checks whether this explanation is consistent with the requirement, calibrating the link's confidence. Like our work, R2Code takes a requirement and the entire codebase as input. However, its filtering step resembles RAG-based approaches such as LiSSA: code files that are essential for understanding how a requirement is implemented may be discarded early. Moreover, errors introduced at each step propagate to the next without any opportunity for correction, and the final self-verification only adjusts confidence scores; it cannot revise predicted links.

In contrast, our first approach lets the LLM reason directly over code files rather than over intermediate summaries and semantic layers, and it does not filter out candidates in advance. Furthermore, we employ a self-correction loop in which a generator proposes trace links, a critic examines them, and an author revises them based on the critic's feedback. Errors in proposed links can thus be corrected before they reach the final output, although links missed by the generator cannot be recovered.
Our second approach uses a multi-agent debate (MAD) framework~\cite{oriol2025}, in which agents argue opposing viewpoints before a judge decides. The framework has been shown to effectively classify requirements as functional or non-functional, but it has not been applied to traceability. In our work, we adapt MAD to requirements-to-code traceability. 

\section{Conclusion} \label{sec:conclusion}

In this paper, we have revisited requirements-to-code traceability in light of the large context windows offered by modern LLMs. Existing approaches rely on a retrieval-and-classification pipeline, so they cannot recover relevant code files that the retrieval step misses; on our benchmarks, retrieval alone caps achievable recall at 59.5\% on average. To address this bottleneck, we have proposed three approaches: \tracer, a single-agent approach that reasons over the entire code repository in one pass, and two multi-agent approaches based on self-correction (\corrector) and multi-agent debate (\debater).

Our evaluation on three benchmark datasets shows that repository-level reasoning substantially improves the coverage of true trace links. With a recall-oriented prompt, \tracer achieves the highest average recall (80.9\%), \SI{38.6}{\pp} above the retrieval-based state-of-the-art, and exceeds the retriever's recall ceiling on every dataset. This gain comes at a significant cost in precision (17.8\% vs.\ 43.6\%), increasing the share of candidate links an analyst must inspect from 1.5--4.7\% to 7.2--17.0\%. Multi-agent validation recovers part of this precision. \corrector removes many false positives but also discards true links and cannot recover links missed in the initial pass. \debater achieves the highest F2 scores among all approaches (50.8\% with the conservative prompt and 49.5\% with the recall-oriented prompt). The latter configuration offers the best balance between validation effort and coverage of true links: it reaches 71.0\% recall, improves F2 by 8.0 and 3.2 percentage points over the state-of-the-art and \tracer, respectively, and reduces the number of links to validate by about 30\% relative to \tracer on eTour and iTrust, though with a larger recall loss on Dronology. Across all approaches, the choice of prompt shifts the precision-recall balance as strongly as the choice of architecture.

Our qualitative analysis shows that the effectiveness of each approach depends on where traceability evidence resides. Retrieval-based approaches are primarily constrained by the quality of requirements, whereas approaches that reason over the full codebase are influenced by code quality, such as documentation and naming conventions. \tracer excels when trace evidence is explicit, whereas \debater is more effective when evidence is indirect, conflicting, or distributed across the implementation. These findings provide practical guidance on which approach to choose depending on the characteristics of the artifacts and the available validation budget.

In future work, we plan to evaluate our approaches on larger industrial repositories, including ones that exceed the context window, across multiple models and programming languages, and to investigate how they can be integrated into development workflows involving AI coding agents.

\section*{Acknowledgment}
This research was funded in part by the Luxembourg National
Research Fund (FNR), under grant number  
C23/IS/17958091/PLAITO. Lionel Briand was funded by the Discovery
Grant program of the Natural Sciences and Engineering Research Council of Canada (NSERC) and the Research Ireland grant 13/RC/209.

\bibliographystyle{IEEEtranDOI}
\bibliography{main}
\end{document}